\documentclass[fleqn,usenatbib]{mnras}

\usepackage{newtxtext,newtxmath}
\usepackage[T1]{fontenc}

\DeclareRobustCommand{\VAN}[3]{#2}
\let\VANthebibliography\thebibliography
\def\thebibliography{\DeclareRobustCommand{\VAN}[3]{##3}\VANthebibliography}

\usepackage{graphicx}	% Including figure files
\usepackage{amsmath}	% Advanced maths commands
\usepackage{multirow}
\usepackage{pdflscape}
\usepackage{subcaption}
\usepackage{xcolor}

\title[The star formation properties of HI galaxies]{MIGHTEE-HI: The star-forming properties of H{\sc i} selected galaxies}

\author[M. N. Tudorache et al.]{Madalina N. Tudorache,$^{1}$\thanks{E-mail: madalina.tudorache@physics.ox.ac.uk}
M. J. Jarvis,$^{1, 2}$
A. A. Ponomareva, $^{1, 5}$
I. Heywood, $^{1, 3, 4}$
N. Maddox, $^{6}$
\newauthor
B. S. Frank, $^{7, 8, 4, 9}$
M. Baes, $^{10}$
R. Dav{\'e}, $^{11, 2}$
S. L. Jung, $^{1}$
M. Maksymowicz-Maciata, $^{6}$
H. Pan, $^{1}$
\newauthor
K. Spekkens $^{12}$
\\
$^{1}$Astrophysics, Department of Physics, University of Oxford, Keble Road, Oxford OX1 3RH, UK\\
$^{2}$Department of Physics and Astronomy, University of the Western Cape, Robert Sobukwe Road, 7535 Bellville, Cape Town, South Africa\\
$^{3}$Department of Physics and Electronics, Rhodes University, PO Box 94, Makhanda, 6140, South Africa \\
$^{4}$South African Radio Astronomy Observatory, 2 Fir Street, Black River Park, Observatory, Cape Town 7925, South Africa\\
$^{5}$Centre for Astrophysics Research, School of Physics, Astronomy and Mathematics, University of Hertfordshire, College Lane, Hatfield AL10 9AB, UK\\
$^{6}$School of Physics, H.H. Wills Physics Laboratory, Tyndall Avenue, University of Bristol, Bristol, BS8 1TL, UK \\
$^{7}$STFC UK Astronomy Technology Centre, Royal Observatory, Edinburgh, Blackford Hill, Edinburgh, EH9 3HJ \\
$^{8}$Department of Astronomy, University of Cape Town, Private Bag X3, Rondebosch 7701, South Africa \\
$^{9}$The Inter-University Institute for Data Intensive Astronomy (IDIA), and University of Cape Town, Private Bag X3, Rondebosch, 7701, South Africa \\
$^{10}$Sterrenkundig Observatorium, Universiteit Gent, Krijgslaan 281 S9, B-9000 Gent, Belgium \\
$^{11}$Institute for Astronomy, University of Edinburgh, Royal Observatory, Blackford Hill, Edinburgh EH9 3HJ, UK\\
$^{12}$Department of Physics, Engineering Physics \& Astronomy, Queen’s University, Kingston, ON K7L 3N6, Canada\\
}

\date{Accepted XXX. Received YYY; in original form ZZZ}

\pubyear{2024}

\begin{document}
\label{firstpage}
\pagerange{\pageref{firstpage}--\pageref{lastpage}}
\maketitle

\begin{abstract}
The interplay between atomic gas, the star-formation history of a galaxy and its environment are intrinsically linked, and we need to decouple these dependencies to understand their role in galaxy formation and evolution.
In this paper, we analyse the star formation histories (SFHs) of $187$ galaxies from the MIGHTEE-H{\sc{i}} Survey Early Science Release data, focusing on the relationships between H{\sc{i}} properties and star formation. A strong correlation emerges between a galaxy’s H{\sc{i}}-to-stellar mass ratio and the time of formation, alongside an inverse correlation between stellar mass and time of formation, regardless of the inferred SFH. Additionally, galaxies with lower stellar masses and higher H{\sc{i}}-to-stellar mass ratios exhibit longer gas depletion times compared to more massive galaxies, which appear to have depleted their gas and formed stars more efficiently. This suggests that smaller, gas-rich galaxies have higher depletion times due to shallower potential wells and less efficient star formation. 
%The atomic gas depletion time ranges from $5$ to $100$~Gyr, indicating that variations in the star formation main sequence are likely driven by long-term environmental effects, rather than short-term fluctuations. 
Furthermore, we explore the connection between spin-filament alignment and H{\sc{i}} content.
%noting that previous studies have shown that H{\sc{i}}-poor galaxies are more aligned with filaments, experiencing minimal angular momentum disruption, while H{\sc{i}}-rich galaxies are often misaligned, potentially due to mergers. 
We find no significant correlation between peak star formation activity and proximity to filaments. However, we do find that the two galaxies in our sample within 1~Mpc of a filament have very low gas-depletion timescales and have their spin axis misaligned with the filament, suggestive of a link between the galaxy properties and proximity to a filament.

\end{abstract}


%%%%%%%%%%%%%%%%%%%%%%%%%%%%%%%%%%%%%%%%%%%%%%%%%%

%%%%%%%%%%%%%%%%% BODY OF PAPER %%%%%%%%%%%%%%%%%%

\section{Introduction}
\label{sec:introduction}

There is a bimodality in the population of galaxies: there are blue, younger, star-forming galaxies and red, quenched galaxies \citep{Kennicutt_1998, Strateva_2001, Baldry_2004, Balogh_2004b, McKee_2007, Kennicutt_2012}. The processes involved in how galaxies stop forming stars are complex and there has been a lot of effort devoted to understand what processes lead to the quenching of galaxies, leading them to move from the blue cloud to the red sequence. From feedback processes - due to both supernovae \citep{Silk_2012, Hopkins_2014} and active galactic nuclei \citep{DiMatteo_2005, King_2005, Croton_2006, Fabian_2012} - to mergers \citep{Barnes_1991, Lacey_1993} and environmental processes, there are many effects that need to be taken into account to understand how galaxies depart from the star-forming main sequence (MS), which links the stellar mass of a galaxy and its star formation activity \citep{Noeske_2007, Whitaker_2012, johnston2015, Popesso_2019a, Popesso_2019b, Leslie_2020, Thorne_2021, Fraser-McKelvie_2021}. Whilst the overall shape of the MS is generally agreed on, such that it increases  up to a knee mass and then flattens before the star formation rate drops precipitously \citep{Lilly_2013, Whitaker_2014, Popesso_2019b}, its scatter is less understood \citep{Whitaker_2015, Matthee_2019}. 

The star formation history (SFH) encodes the temporal narrative of a galaxy's star formation activity. Inferring SFHs involves deciphering the distribution of stellar ages, shedding light on the intensity and duration of past star-forming epochs, potentially providing insights into the physical mechanisms that terminate star formation \citep[e.g.][]{Schreiber_2018}. Similarly, the SFHs of star-forming galaxies can give us information about the assembly of their stellar masses \citep[e.g.][]{Leitner_2012}. The study of stellar populations through colour-magnitude diagrams, spectroscopy, and sophisticated modelling techniques can contribute to disentangling the complex interplay of factors influencing star formation, such as gas availability, environmental conditions, and feedback processes \citep{Ocvirk_2006, Dye_2008, Leja_2017, Leja_2019, Tacchella_2022}.

Tying everything together, there are several prevailing theories that aim to explain the scatter in the MS, related to the SFH of a galaxy. Variations in SFHs, driven by episodic or bursty star formation, lead to deviations from the star-forming MS as galaxies experience different phases of heightened or suppressed star formation \citep{Finlator_2008, Lilly_2013, Dekel_2014, Tacchella_2016}. These fluctuations are closely tied to the dynamics of gas flows, where inflows of cold gas replenish the fuel for star formation, causing temporary surges in star formation rates that push galaxies above the star-forming MS. On the other hand, outflows driven by feedback mechanisms like stellar winds and active galactic nuclei (AGN) can deplete the gas reservoir, reducing SFRs and causing galaxies to fall below the sequence. The balance between these gas inflows and outflows, along with the efficiency of gas conversion into stars, contributes to the diversity of star formation histories and the resulting scatter around the star-forming MS \citep{Abramson_2015, Munoz_2015}. However, there are cases in which the SFH variations do not necessarily have any dramatic bursts or quenching events, such that their offset from the MS is a long timescale effect \citep{Rodriguez-Puebla_2016, Matthee_2019}. These galaxies might be in environments that slowly strip away their gas, such as in galaxy clusters where processes like ram pressure stripping or strangulation gradually remove the cold gas needed for star formation. Then, over time, this slow decline in gas content leads to a reduction in star formation rates, causing these galaxies to drift below the star-forming MS while maintaining a stable, but uneventful SFH \citep{Peng_2010, Behroozi_2013}. 
Hence, if the MS scatter arises from to short-term fluctuations, star-forming galaxies with similar masses mostly grew self-similarly. However, if the scatter in the MS arises due to longer-term fluctuations, then star-forming galaxies with similar mass may not have evolved in a similar way and the key physical mechanisms lie in the processes that diversify the SFHs.

The environment in which these galaxies live will clearly play a role in how their gas behaves. The effect of the local environment on galaxies (i.e. clusters) has been well-researched \citep[e.g.][]{Dressler_1980, Davis_Geller_1976, Balogh_2004b, Peng_2012, Robotham_2013, Treyer_2018, Davies_2019}. However, the effects of the large-scale cosmic web \citep{bond_1996} and its components (filaments, walls and voids), in particular the filamentary structures are not as well understood. In the context of H{\sc i}, \citet{kleiner_2017} and \citet{crone-odekon_2018} have investigated the link between H{\sc i} in galaxies and the large-scale structures, with different results regarding the correlation between position of the galaxy with respect to the cosmic filaments and its H{\sc i} content, and how they could be fuelled by them. The former found that massive galaxies ($M_{\ast} > 10^{11}$\,M$_{\odot}$) have increased H{\sc i}-to-stellar-mass ratios closer to filaments, implying that galaxies replenish some of their gas from the intra-filamentary medium. However, the latter argue that lower-mass galaxies ($M_{\ast} < 10^{10.5}$\,M$_{\odot}$) show an H{\sc i} deficiency closer to filament spines due to a cut-off from their gas supply by the filaments. There are other properties of H{\sc i} selected galaxies that can be investigated as a function of distance to large-scale structures (in this case, filaments), such as the angular momentum of a galaxy. The overall picture of the relationship between the spin vector of a galaxy and its alignment with the filaments of the cosmic web in which it may reside appears complicated. There have been a few studies using H{\sc i} galaxies which proposed a more cohesive picture of this spin-filament alignment. For example, a potential spin transition threshold in H{\sc i} mass at $M_{\text{H{\sc{I}}}} = 10^{9.5}$M$_{\odot}$ is identified by \citet{Kraljic_2020} through analysis of the SIMBA simulation \citep{simba-sim}. Meanwhile, \citet{bluebird_2019}, utilising the COSMOS H{\sc i} Large Extragalactic Survey \citep[CHILES;][]{Dodson_2022}, observes that galaxies in their H{\sc i}-selected sample often exhibit alignment with the cosmic web. However, their investigation does not reveal a substantial mass transition between galaxies with aligned and misaligned spins. \cite{tudorache2022}, using a much larger sample from the MIGHTEE-H{\sc i} \citep{Maddox_2021} Early Release data, do not find a link between the H{\sc i} mass of galaxies and their spin-filament alignment. However, they find that galaxies which are misaligned have higher H{\sc i}-to-stellar mass ratios. One possible explanation for this effect is the occurrence of gas-rich mergers in these galaxies, which, in turn, could cause the mis-alignment in the spin-filament angle, whilst increasing the mass of the H{\sc i} gas in the galaxies. As mergers have been associated with the presence of starbursts in galaxies in both observations \citep[e.g.][]{Lin_2010, Chou_2013, Robotham_2014} and simulations \citep[e.g.][]{Renaud_2014, Moreno_2019, Cenci_2024}, one way to investigate this theory is by measuring the SFHs of these galaxies and checking for any bursts.
\begin{figure}
    \centering
	\includegraphics[width=\columnwidth]{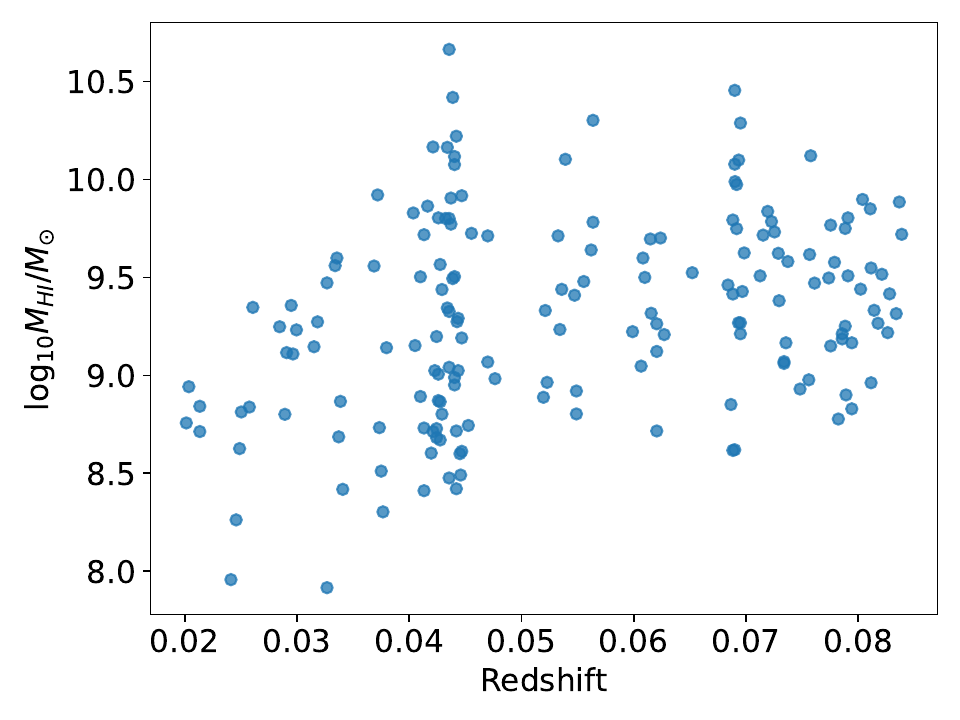}
    \caption{Atomic gas mass $M_{{\text{H\sc i}}}$ as a function of redshift for the MIGHTEE-H{\sc i} Early Release Science catalogue used in this work.}
    \label{fig:mhi-z}
\end{figure}

For this study, we use a H{\sc i} galaxy sample provided by the MeerKAT International GigaHertz Tiered Extragalactic Exploration (MIGHTEE,  \citealt{mightee}) Early Science release to investigate the SFHs of an H{\sc i} selection sample and how they link to their large-scale environment. We assume $\Lambda$CDM cosmology with $H_0 = 70$ km\,s$^{-1}$\,Mpc$^{-1}$ and $\Omega_{\rm M} = 0.3$ and $\Omega_\Lambda = 0.7$. The structure of this paper is organised as follows. Section \ref{sec:data} introduces the data used in this study and the process of fitting of spectral energy distributions to infer the galaxy properties. Section~\ref{sec:results} discusses the results obtained. The summary and conclusions are presented in Section~\ref{sec:conclusions}.

\begin{figure*}
  	\includegraphics[width=1\textwidth]{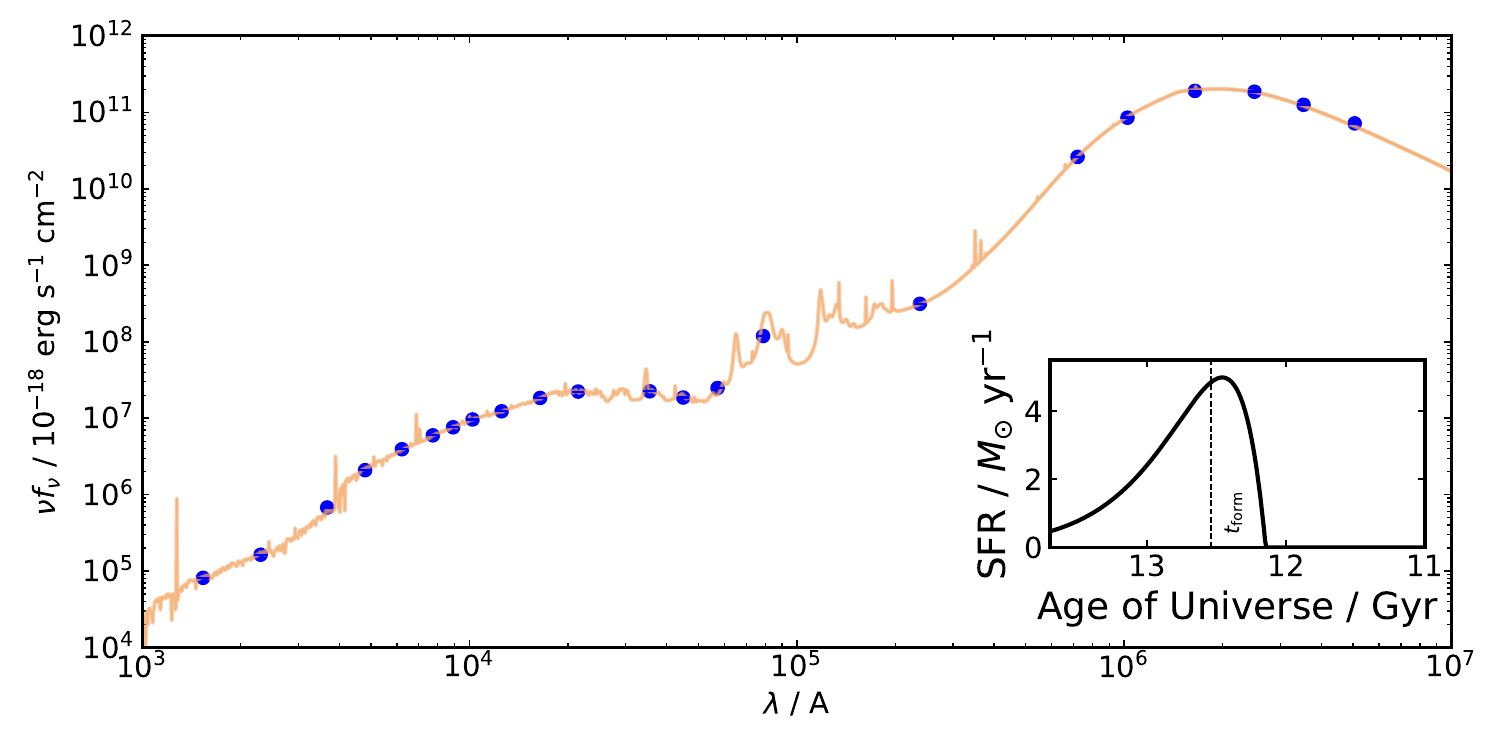}
    \caption{An example of the photometric output of one of the MIGHTEE-H{\sc i} Early Science galaxies using the available photometric filters, fitted with an exponentially-delayed SFH (presented as an inset in the bottom right corner, along with a dashed vertical line showing $t_{\text{form}}$). This is a galaxy with SFR = 0.54 M$_{\sun}/\text{yr}$, $M_{\ast} = 10^{9.38}$~M$_{\sun}$ and $M_{\text{H\sc i}} = 10^{8.98}$~M$_{\sun}$.}
    \label{fig:photo-sed_20}
\end{figure*}

\section{Data}\label{sec:data}
\begin{table}
\centering
\caption{Short summary of the MIGHTEE-H{\sc i} Early Science
  data products used in this paper.}
\label{tab:mightee}
\begin{tabular}{p{3.5cm} p{3.5cm}}
\hline\hline
\multirow{2}{*}{Area covered} & $\sim 1$\,deg$^2$ COSMOS field \\
                  &  $ \sim 3$\,deg$^2$ XMM-LSS field  \\
Frequency range   &  $ 1320 - 1410 $\,MHz \\
Redshift range   &  $ 0.004 - 0.084 $ \\
Channel width     &  $209$\,kHz \\
%Pixel size        &  $2$\,arcsec \\
Median $\text{H}_{\text{I}}$ channel rms noise  &  $85$ $\mu$Jy \,beam$^{-1}$\\
$N_{\text{HI}}$ sensitivity ($3 \sigma$)   & $1.6 \times 10^{20}$ \,cm$^{-2}$ (per channel) \\ 
\multirow{2}{*}{Synthesised beam} & $14.5" x 11"$ COSMOS field \\
                                 & $12" x 10"$ XMM-LSS field \\

\hline \hline
\end{tabular}
\end{table}

\subsection{The MIGHTEE survey}
\label{subsec:mightee}

The MIGHTEE survey is one of eight Large Survey Projects using MeerKAT \citep[][]{meerkat}. MeerKAT comprises a configuration of 64 offset-Gregorian dishes, each featuring a main reflector with a 13.5 m diameter and a sub-reflector with a $3.8$~m diameter. The receivers for MeerKAT operate across three bands: UHF–band ($580 < \nu < 1015 $ MHz), L--band ($900 < \nu < 1670$ MHz), and S–band ($1750 < \nu < 3500$ MHz), all capable of collecting data in spectral-line mode. The MIGHTEE survey focuses on three main aspects: radio continuum \citep{Heywood2021, Hale2024}, polarisation \citep{Taylor2024}, and spectral line \citep{Heywood_2024}. 

MIGHTEE-H{\sc i} \citep{Maddox_2021} constitutes the H{\sc i} emission project within the MIGHTEE survey. 
Data products, released as part of the Early Science phase, were provided using the \texttt{ProcessMeerKAT} calibration pipeline \citep{Collier_2021}. This pipeline, based on \texttt{CASA}\footnote{\url{http://casa.nrao.edu}} \citep{casa}, operates in a parallelised manner and follows standard calibration routines such as flagging, delay, bandpass, and complex gain calibration. Spectral-line imaging is performed using \texttt{CASA}'s \texttt{TCLEAN} task. Continuum subtraction was executed in two domains. Visibility domain subtraction utilised standard \texttt{CASA} routines, namely \texttt{UVSUB} and \texttt{UVCONTSUB}. Subsequently, image plane-based continuum subtraction was implemented through per-pixel median filtering, applied to the resulting data cubes to mitigate the impact of direction-dependent artefacts.  The summary of the data utilised in this paper is provided in Table \ref{tab:mightee}. In this study, we specifically utilise the spectral line data from the L--band Early Science \citep[see ][ for source catalogues]{Ponomareva2023}, employing $4096$ channels with a channel width of $209$~kHz, corresponding to a velocity resolution of $44$~km s$^{-1}$ at redshift $z=0$. These observations cover two of the four MIGHTEE fields, namely COSMOS and XMM-LSS, where we have excellent multi-wavelength ancillary data to determine the properties of H{\sc i}-selected galaxies from MIGHTEE.

%\begin{figure*}
%\begin{subfigure}[b]{\columnwidth}
%  \centering
  % include first image
%  \captionsetup{justification=centering}
%  \includegraphics[width=1\linewidth]{figures/crossmatch_COSMOS.pdf} 
%  \phantomcaption
%  \label{subfig:cosmos-crossmatch}
%\end{subfigure}
%\begin{subfigure}[b]{\columnwidth}
%  \centering
  % include second image
%  \captionsetup{justification=centering}
%  \includegraphics[width=1\linewidth]{figures/crossmatch_XMM.pdf}
%  \phantomcaption
%  \label{subfig:xmm}
%\end{subfigure}
%\caption{The sources from the MIGHTEE-H{\sc i} Early Release catalogue (blue points) and their crossmatches from the DEVILS catalogue (pink points) for the COSMOS ({\it left}) and XMM-LSS ({\it right}) fields.}
%  \label{fig:crossmatch}
%\end{figure*}

There are $276$ galaxies in the full Early Science H{\sc i} catalogue.
%We used the ancillary data extracted by the MIGHTEE-H{\sc i} team for the $ugrizYJHK_s$ photometry as detailed in \cite{Maddox_2021}. %Furthermore, we crossmatched these galaxies with the Herschel Extragalactic Legacy Project \citep[HELP;][]{Shirley_2019} in order to obtain mid-to-far infrared data. As some of the galaxies do not have full photometry available, and because we are restricted to the $0.02 < z < 0.09$ redshift interval for the filaments, our sample was reduced to $219$ galaxies. Within this sample, we retained $66$ out of the $77$ the galaxies used in \cite{tudorache2022} to study the spin-filament alignment of the galaxies with the filaments of the cosmic web, due to the same reason (lack of photometry in some bands) as for the full sample.
We cross-matched these galaxies with the Deep Extragalactic VIsible Legacy Survey \citep[DEVILS;][]{devils} photometric catalogue \citep{devils-photom}, in order to obtain ultraviolet through to mid- and far-infrared data measured in a consistent way using ProFOUND \citep{Robotham2018}. The DEVILS photometric catalogue is derived from the imaging data using the {\em Galaxy Evolution Explorer} \citep[GALEX; ][]{Zamojski2007} for ultraviolet wavelengths, the Canada-France-Hawaii Telescope \citep[CFHT; ][]{Ilbert2006,Capak2007} ($u$-band), HyperSuprimeCam \citep[HSC; ][]{Aihara2019} ($grizy$), Visible-Infrared Survey Telescope for Astronomy \cite[VISTA; ][]{McCracken2012, Jarvis2013} ($YJHK_{s}$), {\em Spitzer Space Telescope} \citep[][]{Lonsdale2003,Sanders2007,Mauduit2012} (mid-infrared, IRAC ch1-ch4 and MIPS $24~\mu m$,  $70~\mu m$), and the {\em Herschel Space Observatory} \citep[][]{Oliver2012} (far-infrared, PACS and SPIRE). This plethora of very deep imaging data from the UV through to the far-infrared allows us to perform galaxy SED fitting to much better accuracy than has been possible for H{\sc i}-selected samples. Importantly, by sampling both the rest frame UV and far-infrared emission, we are sensitive to the total star-formation rate in these galaxies. 

However, some of the galaxies do not have full photometry available due to the MIGHTEE Early Science Release data extending beyond the slightly smaller areal coverage of the optical and near-infrared data, reducing the sample to $187$ galaxies. Figure \ref{fig:mhi-z} shows the distribution of their H{\sc i} mass as a function of redshift.
%Within this sample, we also only retained $51$ out of the $77$ the galaxies used in \cite{tudorache2022} to study the spin-filament alignment of the galaxies with the filaments of the cosmic web, due to the same reason (lack of photometry in some bands) as for the full sample.

\subsection{SED fitting of galaxies}
\label{sec:sedfitting}

There are several different algorithms that can be used for fitting spectral energy distributions (SEDs) of galaxies in order to obtain physical information from photometric data \citep[e.g.][]{lephare2, ProSPECT, Leja_2017}. Known as template fitting, it involves fitting observed photometric data with theoretical templates or model spectra \citep{Bolzonella_2000, lephare2, Walcher_2011, Hunt_2019, Pacifici_2023}.
%The fitting can be done in several ways, by using prior knowledge about the expected properties of the galaxy, such as stellar population models and dust attenuation, to derive probability distributions for the physical parameters. 
%SED fitting techniques have been used to extract valuable information about the age, composition, and physical conditions of galaxies based on their observed photometric data \citep{Bolzonella_2000, lephare2, Walcher_2011}. 

In this paper, we use \textsc{Bagpipes}\footnote{\url{http://bagpipes.readthedocs.io}} \citep{Carnall_2018}, which is a \textsc{Python} tool that uses Bayesian inference to fit model SEDs to measured galaxy SEDs and provide redshifts and galaxy properties using spectroscopic and/or photometric data from the ultraviolet to the microwave regime. 
%Specifically, it is an algorithm based on a nested sampling approach using \textsc{Multinest} \citep{Feroz_2008, Feroz_2009}.
\textsc{Bagpipes} provides a framework for computing both parametric and non-parametric SFHs, such as delta functions, constant, exponentially declining, delayed exponentially declining, log-normal, double-power law or any custom input (see \citealt{Carnall_2018} for all the functional forms). For our sample of galaxies, given that we already have redshifts (from both DEVILS and H{\sc i} measurements), we fix the redshift values and run \textsc{Bagpipes} with the \cite{bruzual2003} stellar population model, which is characterised by a \citet{Chabrier_2003} IMF. For dust, we apply the \cite{Calzetti2000} attenuation law with priors on E(B-V) $ = (0.0, 3.0)$. We also apply the \cite{Cardelli_1989} attenuation law with similar priors. However, we only show the results of the \cite{Calzetti2000} attenuation law, as the trends do not change. The dust emission from the neutral ISM  is modelled as a single-temperature grey-body \citep{Hildebrand_1983}. Similarly, we use uniform priors for all other properties (Table~\ref{table:fitting-priors}). An example of the fit can be seen in Figure~\ref{fig:photo-sed_20}. Examples of fits for other galaxies (as well as corner plots) can be seen in Appendix~\ref{sec:appendix-a}.

\begin{table}
    \centering
    \caption{A description of each of the parameters used for the exponentially-delayed SFH model, as well as the priors used to fit the model. The parameters are (from top to bottom): the age of the galaxy, the SFR e-folding time $\tau$, the stellar mass of the galaxy $M_{\star}$, the metallicity $Z$ in units of solar metallicity, $Z_0$), the dust attenuation coefficient $A_V$, the PAH mass fraction $q_{\text{PAH}}$, the lower limit of starlight intensity distribution $u_{\text{min}}$, the fraction of stars at $u_{\text{min}}$ $\gamma$ and the ionisation parameter, $U$.}
    \label{table:fitting-priors}
\begin{tabular}{|cc|}
\hline
\textbf{Parameter} & \textbf{Prior distribution} \\
\hline \hline
Age & uniform $\in [0.1, 15.0]$ \\
$\tau$ & uniform $\in [0.3, 10.0]$\\
$\log{M_{\ast}}$ & uniform $\in [1.0, 15.0]$\\
$\log{Z}$ & uniform $\in [0.0, 2.5]$\\
A$_{V}$ & uniform $\in [0.0, 3.0]$ \\
q$_{\text{PAH}}$ & uniform $\in [0.1, 4.58]$ \\
u$_{\text{min}}$ & uniform $\in [0.1, 20.0]$ \\
$\gamma$ & uniform $\in [0.0, 0.5]$ \\
$\log_U$ & uniform $\in [-4.0, -1.0] $ \\
\hline \hline
\end{tabular}
\end{table}

In Bayesian inference, the statistical distribution of a set of parameters, $\boldsymbol{\theta}$, for a model $\mathcal{M}$, given some data, $\boldsymbol{d}$, is given by Bayes theorem: 
\begin{equation}
\mathcal{P}(\boldsymbol{\theta} | \boldsymbol{d}, \mathcal{M})=\frac{\mathcal{L}(\boldsymbol{d} | \boldsymbol{\theta}, \mathcal{M}) \Pi(\boldsymbol{\theta} | \mathcal{M})}{\mathcal{Z}(\boldsymbol{d} | \mathcal{M})}
\label{eq:bayes-theorem}
\end{equation}
where $\mathcal{P}(\boldsymbol{\theta} | \boldsymbol{d}, \mathcal{M})$ is the posterior probability, $\mathcal{L}(\boldsymbol{d} | \boldsymbol{\theta}, \mathcal{M})$ is the likelihood, $\Pi(\boldsymbol{\theta} | \mathcal{M})$ represents the priors and  $\mathcal{Z}(\boldsymbol{d} | \mathcal{M})$ is the evidence. 
As \textsc{Bagpipes} uses a Bayesian inference approach to compute the SFHs and the stellar masses, we can also use the evidence ($\log{\mathcal{Z}(\boldsymbol{d} | \mathcal{M}_i)}$) for each model $i$ and for each galaxy. We can then use the Bayes factor and the Jeffreys scale \citep{jeffreys_1998} to verify which model is preferred accounting for model complexity. We define the Bayes' factor as:

\begin{equation}
    \log{B_{01}} \equiv \log{\mathcal{Z}(\boldsymbol{d} | \mathcal{M}_0)} - \log{\mathcal{Z}(\boldsymbol{d} | \mathcal{M}_1)},
    \label{eq:log-bayes-factor}
\end{equation}
where $\mathcal{M}_0$ and $\mathcal{M}_1$ are two models which we are comparing. We take $\log{B_{01}} < 1$ as "not significant", $ 1 < \log{B_{01}} < 2.5$ as "significant", $ 2.5 < \log{B_{01}} < 5$ as "strong", and $\log{B_{01}} > 5$ as "decisive" \citep{jeffreys_1998} as a way to determine the preferred model given the data and number of free parameters.

\section{Results and Discussion}
\label{sec:results}

\subsection{Galaxy SED models}
\label{subsec:sed-comp}

As described in Section~\ref{sec:sedfitting}, \textsc{Bagpipes} provides many models that can be fit to a galaxy in order to infer its SFH. We use the six available parametric SFH models for our galaxy sample: the exponential SFH, the constant SFH, the log-normal SFH, the delayed SFH, the burst SFH and the double-power law SFH. 
%The MIGHTEE-H{\sc i} full Early Science release catalogue ancillary data also provides stellar masses. The SED fitting code \textsc{LePhare} \citep{lephare1, lephare2} was used to derive the stellar mass. The uncertainty in stellar mass for each galaxy that was adopted is $ \sim 0.1$\,dex \citep{adams_2021}. However, we choose to use the stellar mass and SFRs computed from \textsc{Bagpipes}. 
To compare the SFH models from \textsc{Bagpipes}, we use the Bayesian framework described in Section \ref{sec:sedfitting}. We calculate the Bayes' factor for each galaxy and for each model and proceed to draw a comparison between them. When comparing the SFH models for each galaxy, we find that the broadband data is not sufficient to differentiate between the exponential SFH, the lognormal SFH and the exponentially-delayed SFH model. Therefore, for the rest of this paper we will only show the full results for the exponentially-delayed SFH model. For reference, the exponentially-delayed SFH model is defined as:

\begin{equation}
\operatorname{SFR}(t) \propto \begin{cases}e^{-\frac{t-T_0}{\tau}} & t>T_0 \\ 0 & t<T_0\end{cases}
\end{equation}
where $\tau$ is the SFR e-folding time and $T_0$ is the cosmic time.

% \begin{figure*}
%   	\includegraphics[width=1\textwidth]{figures/184_sfh.pdf}
%     \caption{An example of an exponentially delayed SFHs of one of the MIGHTEE-H{\sc i} Early Science galaxies using the available photometric filters.}
%     \label{fig:delayed-sfh}
% \end{figure*}

\begin{figure*}
  	\includegraphics[width=1\textwidth]{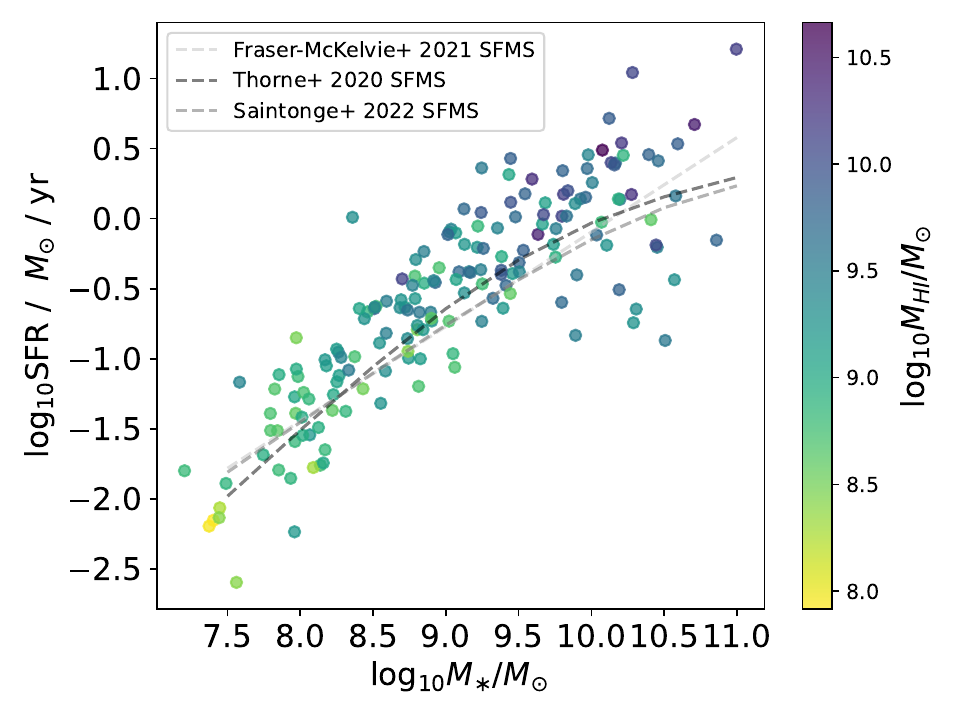}
    \caption{The stellar mass versus star-formation for our H{\sc i}-selected galaxies, with the properties derived from the best-fitting model using {\sc Bagpipes}. The H{\sc i} mass is denoted by the colour bar. It is clear that the H{\sc i}-selection results in a sample of galaxies that lie on the main-sequence of star-forming galaxies, albeit with the often-seen turnover in star formation for the more massive galaxies in the sample showing that quenched galaxies may still retain a significant H{\sc i} reservoir. The three dashed lines are different SF-MS fits: the black line is the SF-MS as presented in \citet{Thorne_2021}, the light gray line is the linear SF-MS as presented in \citet{Fraser-McKelvie_2021} and the dark grey line is the SF-MS as presented in \citet{Saintonge_2022}. All the SF-MS relations presented  are computed from optically-selected samples, and not an H{\sc i}-selected sample.}
    \label{fig:sfms}
\end{figure*}

\subsection{The stellar properties of H{\sc i}-selected galaxies}
\label{subsec:hi-props}

The processes involved in the conversion of H{\sc i} to stars is complex and not well understood \citep{Maddox_2021}. In the following, we investigate links between the H{\sc i}-selected galaxy sample and several parameters which describe star formation.
%such as the peak time of star formation, $t(z_{\text{form}})$ or the atomic gas depletion time, $t_{\text{dep}}$, as inferred from \textsc{Bagpipes}. 
%Specifically, we are running these tests since H{\sc i} serves as the raw material for the build-up of stellar mass in galaxies; however t
It should be noted that all the results presented below are in the context of an H{\sc i}-selected sample, which provides unique insights into the most gas-rich systems. However, the fact that a galaxy must have significant amounts of H{\sc i} in order to be detected means that the effect of low-mass H{\sc i} galaxies on these relations will not be observed. This then also leads to being biased towards a sample formed of mostly spiral and irregular galaxies. Furthermore, galaxies with higher H{\sc i} content are often actively forming stars, leading to a bias towards younger stellar populations, and galaxies on the main sequence. However, we are sensitive to low-luminosity dwarf galaxies, which tend to be gas-dominated \citep{Geha_2006} and not biased towards high surface brightness galaxies. Conversely to many optically-selected samples used to investigate the dependence on H{\sc i}, we are not significantly constrained by the luminosity of the stellar population, stellar mass, or dust extinction \citep{Helmboldt_2004, Rosenberg_2005, Martin_2012, Huang_2012a}, as the multi-wavelength data is very deep for the H{\sc i}-selected galaxies at these very low ($z<0.08$) redshifts \citep[e.g. see][for more information]{adams_2021,devils-photom}. In Figure~\ref{fig:sfms}, we show the stellar mass versus star-formation rate for our H{\sc i}-selected galaxies. This demonstrates that our galaxies are predominantly the H{\sc i}-rich galaxies lying along or above the main sequence of star formation. However, we also find significant H{\sc i} gas reservoirs ($M_{\text{H\sc i}} > 10^{9} $~M$_{\sun}$) in the more massive galaxies that are beginning to turn off the main sequence. %or in one case is completely quenched with a specific star-formation rate $\text{sSFR} \sim 10^{-13}$\,yr$^{-1}$ (see Figure~\ref{fig:galaxy_111}).

In the following sections, we discuss the relationship between the H{\sc i} and the derived properties from the SED fitting with {\sc Bagpipes}. Full details of the derived parameters are given in Table~\ref{table:data} alongside the redshift from the H{\sc i} line.

\begin{table*}
\centering
\caption{An excerpt of fitted parameters for the exponentially-delayed model on the MIGHTEE-H{\sc i} galaxies, as well as the H{\sc i} mass. Full table is available as supplementary material.}
\label{table:data}
\begin{tabular}{crrrrrrrrr}
\hline\hline
ID & RA & Dec & z & $\log_{10}M_{\text{H{\sc i}}}/\text{M}_{\sun}$ & $\log_{10}{M_{\ast}/M_{\sun}}$ & SFR & sSFR & $t_{\text{form}}$ & $t_{\text{dep}}$ \\
\hline\hline
0& $150.4742$ & $2.4137$ & $0.0068$ & $7.69$ & $7.29_{-0.001}^{+0.001}$ & $0.007_{-0.001}^{+0.001}$ & $-9.456_{-0.001}^{+0.001}$ & $12.491_{-0.001}^{+0.001}$ & $9.86_{-0.001}^{+0.001} $\\
1& $150.613$ & $2.1668$ & $0.0059$ & $7.31$ & $6.912_{-0.001}^{+0.001}$ & $0.004_{-0.001}^{+0.001}$ & $-9.269_{-0.001}^{+0.001}$ & $12.522_{-0.001}^{+0.001}$ & $9.666_{-0.001}^{+0.001} $\\
2& $150.1253$ & $2.1496$ & $0.0041$ & $6.69$ & $6.155_{-0.0}^{+0.0}$ & $0.001_{-0.001}^{+0.001}$ & $-9.454_{-0.001}^{+0.001}$ & $12.59_{-0.001}^{+0.001}$ & $9.994_{-0.001}^{-0.001} $\\
3& $149.6951$ & $2.3475$ & $0.0058$ & $8.57$ & $7.342_{-0.001}^{+0.001}$ & $0.005_{-0.001}^{+0.001}$ & $-9.652_{-0.003}^{+0.003}$ & $12.433_{-0.002}^{+0.002}$ & $10.884_{-0.003}^{+0.003} $\\
4& $149.8662$ & $2.007$ & $0.013$ & $8.51$ & $7.425_{-0.003}^{+0.003}$ & $0.016_{-0.001}^{+0.001}$ & $-9.232_{-0.004}^{+0.004}$ & $11.823_{-0.025}^{+0.020}$ & $10.313_{-0.001}^{+0.001} $\\
5& $150.0243$ & $1.911$ & $0.0062$ & $7.66$ & $7.47_{-0.002}^{+0.002}$ & $0.002_{-0.001}^{+0.001}$ & $-10.084_{-0.003}^{+0.002}$ & $11.877_{-0.013}^{+0.013}$ & $10.275_{-0.001}^{+0.002} $\\
6& $150.5946$ & $2.4223$ & $0.0213$ & $8.84$ & $7.963_{-0.001}^{+0.001}$ & $0.026_{-0.001}^{+0.001}$ & $-9.554_{-0.001}^{+0.001}$ & $12.025_{-0.002}^{+0.002}$ & $10.433_{-0.001}^{+0.001} $\\
7& $149.812$ & $2.192$ & $0.0246$ & $8.26$ & $7.447_{-0.001}^{+0.001}$ & $0.009_{-0.001}^{+0.001}$ & $-9.51_{-0.002}^{+0.002}$ & $12.066_{-0.001}^{+0.001}$ & $10.325_{-0.001}^{+0.003} $\\
8& $150.3455$ & $1.7935$ & $0.0249$ & $8.63$ & $7.843_{-0.0}^{+0.0}$ & $0.031_{-0.001}^{+0.001}$ & $-9.356_{-0.001}^{+0.001}$ & $12.241_{-0.001}^{+0.001}$ & $10.139_{-0.001}^{+0.001} $\\
9& $150.5468$ & $2.0216$ & $0.0213$ & $8.71$ & $9.061_{-0.002}^{+0.002}$ & $0.087_{-0.001}^{+0.001}$ & $-10.122_{-0.002}^{+0.002}$ & $10.952_{-0.019}^{+0.017}$ & $9.774_{-0.001}^{+0.001} $\\
10& $149.9643$ & $1.7067$ & $0.025$ & $8.81$ & $7.204_{-0.001}^{+0.001}$ & $0.016_{-0.001}^{+0.001}$ & $-9.003_{-0.001}^{+0.001}$ & $12.569_{-0.001}^{+0.001}$ & $10.612_{-0.001}^{+0.001} $\\
\hline \hline
\hline
\end{tabular}
\end{table*}

\subsubsection{The star-formation history of H{\sc i} selected galaxies}\label{sec:sfh-hi}

First, we investigate the H{\sc i} mass as a function of the time of formation, $t_{\text{form}}$ (measured forwards from the beginning of the Universe), which is defined as:
\begin{equation}
\frac{\int_0^{t_{\text{obs}}} t \operatorname{SFR}(t) \mathrm{d} t}{\int_0^{t_{\text{obs}}} \operatorname{SFR}(t) \mathrm{d} t}=t(z_{\text{form}}),
\end{equation}
where $t_{\text{obs}} \equiv t(z_{\text{obs}})$, which is the redshift at which we observe the galaxies and SFR is the star formation rate \citep{Carnall_2018}.

\begin{figure}
	\includegraphics[width=\columnwidth]{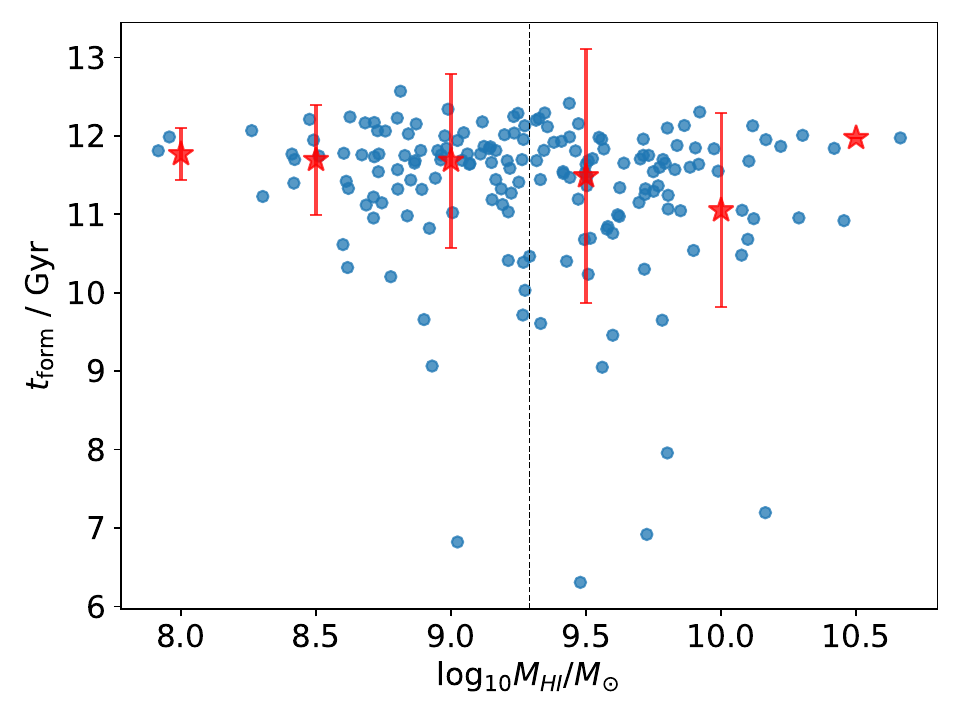}
    \caption{The time of formation $t_{\text{form}}$ as a function of the H{\sc i} mass of a galaxy. The dotted vertical line represents the median $M_{\text{H{\sc i}}}$ value of the sample. The red stars represent the running median of the peak time of star formation $t_{\text{form}}$ as a function of $M_{\text{H{\sc i}}}$. The error bars on the running median are calculated by using the standard deviation of the value in each bin. The vertical dotted line represents the mean H{\sc i} mass of the sample.}
    \label{fig:hi-tform}
\end{figure}

\begin{figure}
	\includegraphics[width=\columnwidth]{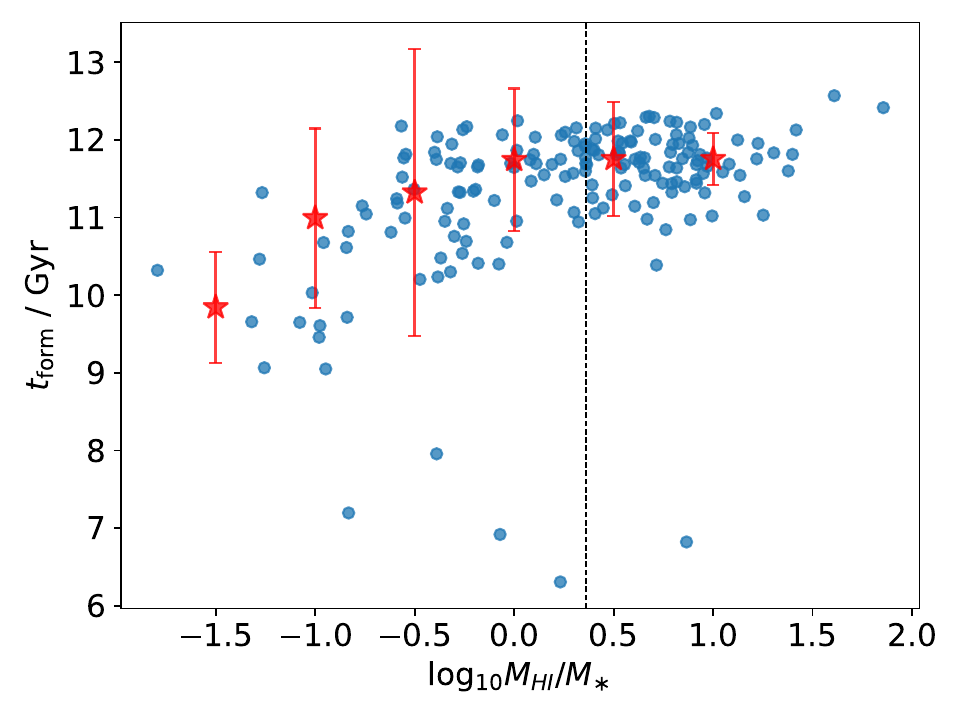}
    \caption{The time of formation $t_{\text{form}}$ as a function of the H{\sc i}-to-stellar mass fraction of a galaxy ($\frac{M_{\text{H{\sc i}}}}{M_{\star}}$). The red stars represent the running median of the peak time of star formation $t_{\text{form}}$ as a function of $M_{\text{H{\sc i}}}/M_{\ast}$. The error bars on the running median are calculated by using the standard deviation of the value in each bin. The vertical dotted line represents the mean H{\sc i}-to-stellar mass ratio of the sample.}
    \label{fig:hi-ms-tform}
\end{figure}

As can be seen in Figure~\ref{fig:hi-tform}, we do not find any relationship between the time since the peak of the star-formation in the galaxy and its H{\sc i} mass. This is confirmed using Kendall's Tau \citep{ktau} and Spearman Rank \citep{src} tests (Table \ref{table:galaxy-prop-stats-delay}). On the other hand, Figure~\ref{fig:hi-ms-tform} shows the time of formation against the H{\sc i}-to-stellar mass ratio and here we see a positive correlation between the two parameters. This is also confirmed by the two correlation tests, shown in Table~\ref{table:galaxy-prop-stats-delay}.  

\begin{figure}
  \centering
  \captionsetup{justification=centering}
  \includegraphics[width=1\linewidth]{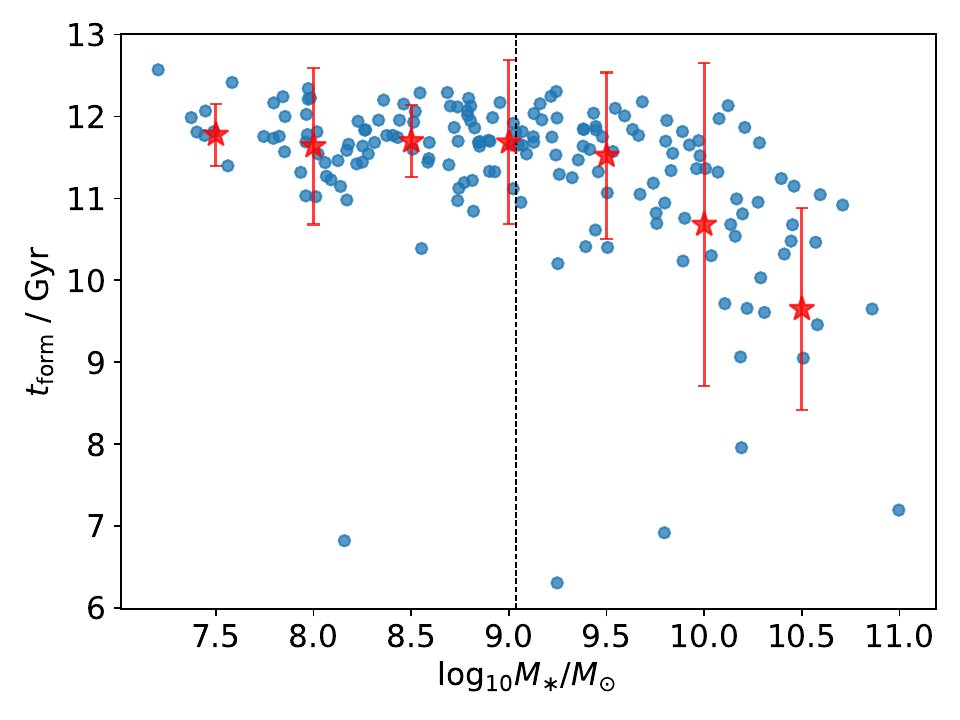} 
\caption{The time of formation $t_{\text{form}}$ as a function the stellar mass $M_{\star}$ of a galaxy. The red stars represent the running median of the peak time of star formation $t_{\text{form}}$ as a function of $M_{\ast}$. The error bars on the running median are calculated by using the standard deviation of the value in each bin. The vertical dotted line represents the mean stellar mass of the sample.}
  \label{fig:stellar-mass-tform}
\end{figure}

As $M_{\text{H{\sc i}}}$ does not show a correlation with the time of formation, $t_{\text{form}}$, but $M_{\text{H{\sc i}}}/M_{\ast}$ does, we also investigate if there is any relationship between the stellar mass of a galaxy $M_{\ast}$ and the time of formation $t_{\text{form}}$. As can be seen in Figure \ref{fig:stellar-mass-tform}, there is an anti-correlation between the time a galaxy reached its peak star-formation and its stellar mass. This is also confirmed by the correlation tests (Table~\ref{table:galaxy-prop-stats-delay}).

Taken together we can infer that older galaxies have less H{\sc i} - which is likely explained by the fact that they have already used the bulk of their cold gas reservoir in order to form stars. This has been also observed in the Survey of Ionisation in Neutral Gas Galaxies \citep[SINGG;][]{Meurer_2006}, in which \cite{Hanish_2006} observed that the most massive galaxies have a larger fraction of their mass locked up in stars compared with H{\sc i}, while the opposite is true for less massive galaxies. Hence, lower mass galaxies are not as evolved because they have converted less of their Inter-Stellar Medium (ISM) into stars than high mass galaxies, so they will have more gas available. The higher H{\sc i}-to-stellar mass ratio in a galaxy is then an indicator of how quickly the SF in a galaxy reached its peak.
%: dwarf galaxies or very late types do not form their stars yet, and that is why the SFH peaks at a later times. 
The stellar mass of a galaxy being anti-correlated with the time when it reached its peak star formation is not entirely surprising, given the result from the H{\sc i}-to-stellar mass ratio. It follows that galaxies with higher stellar masses have likely formed stars earlier in their evolution (as can be seen in Figure \ref{fig:stellar-mass-tform}) and have used up a larger portion of their gas reservoirs. This is essentially a manifestation of cosmic downsizing \citep{Cowie_1996, DeLucia_2006, Thomas_2010} whereby the most massive systems formed earlier.

\subsubsection{The gas depletion timescale of H{\sc i}-selected galaxies}
%We also investigate the links between the atomic gas depletion time of a galaxy $t_{\text{dep}}$ and its properties. By using the computed SFR from the SFHs with the $M_{\text{H{\sc i}}}$ measured from the $21$cm line, the atomic gas depletion time is calculated as:
The gas depletion timescale is defined by the ratio of the gas in a galaxy and the current ongoing star formation. In the case of the H{\sc i} depletion timescale there is obviously an additional link to how quickly the neutral atomic reservoir condenses to form molecular gas from which the stars can form. Nonetheless, it provides a useful indication of how quickly the pristine gas reservoir in the ISM and circumgalactic medium (CGM) is used to fuel star formation.

The atomic H{\sc i} gas depletion timescale is simply defined as:
\begin{equation}
    \centering
    t_{\text{dep}} = \frac{M_{\text{H{\sc i}}}}{\text{SFR}}.
    \label{eq:t-dep}
\end{equation}

\begin{figure*}
\begin{subfigure}[b]{\columnwidth}
  \centering
  % include first image
  \captionsetup{justification=centering}
  \includegraphics[width=1\linewidth]{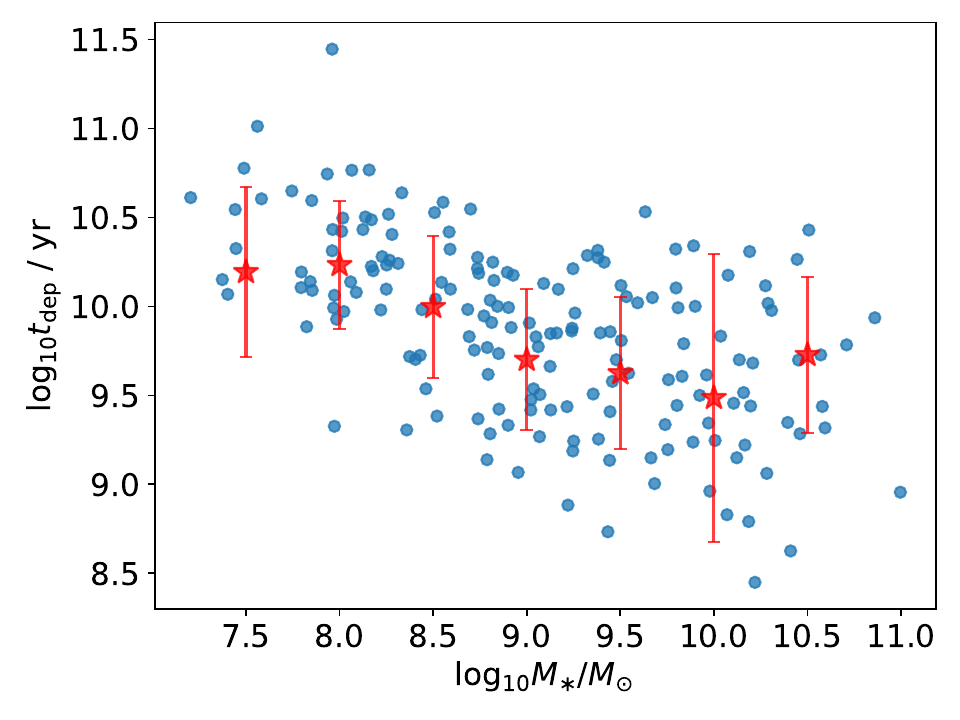} 
  \phantomcaption
  \label{subfig:mhi_sfr_ms}
\end{subfigure}
\begin{subfigure}[b]{\columnwidth}
  \centering
  % include second image
  \captionsetup{justification=centering}
  \includegraphics[width=1\linewidth]{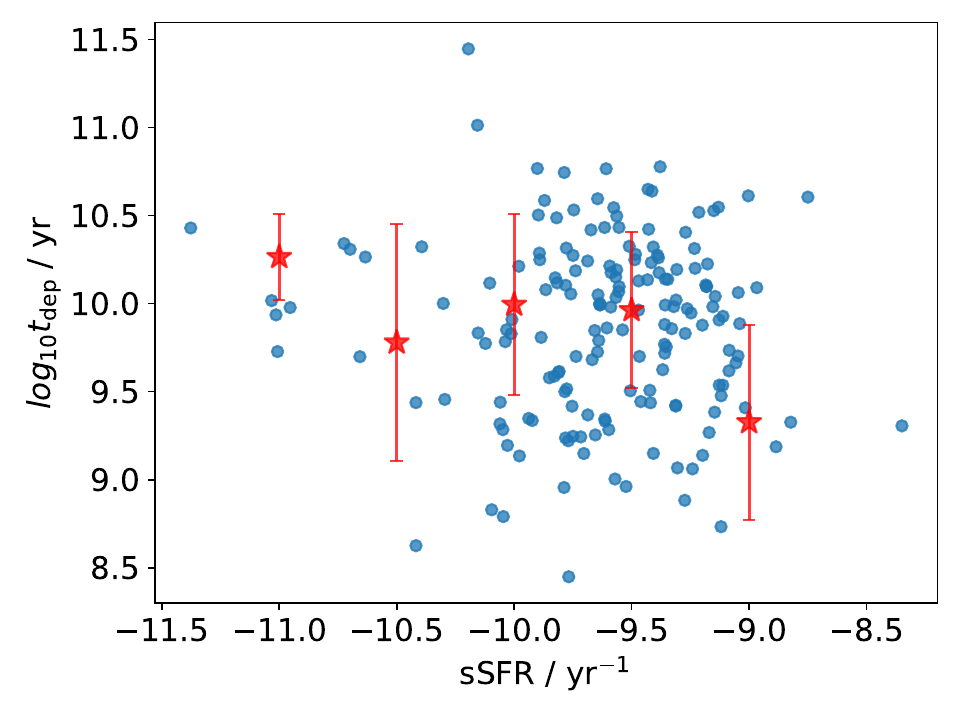}
  \phantomcaption
  \label{subfig:mhi_sfr_ssfr}
\end{subfigure}
\caption{Atomic gas mass depletion $t_{\text{dep}}$ as a function of stellar mass $M_{\ast}$ (left) and sSFR (right). The red stars represent the running median of the atomic gas depletion $t_{\text{dep}}$ as a function of $M_{\ast}$ (left) and sSFR (right) from the observed MIGHTEE-H{\sc i} data. The error bars on the running median are calculated by using the standard deviation of the value in each bin.}
  \label{fig:tdep_scale}
\end{figure*}

The left panel of Figure~\ref{fig:tdep_scale} shows the atomic gas depletion time $t_{\text{dep}}$ as a function of stellar mass for our H{\sc i}-selected galaxies. 
As can be seen, lower stellar mass galaxies tend to have significantly higher gas depletion times than their high-H{\sc i} mass counterparts. This trend is stronger for galaxies with stellar mass $M_{\ast} < 10^{9.5} M_{\odot}$, and flattens out towards higher masses.
%This is also confirmed by the Mann-Whitney U test, where we find a $p-$value of $1.44 \cdot 10^{-7}$. 
This is also confirmed by the correlation tests (Table~\ref{table:galaxy-prop-tdep-stats-delay}). Given that our galaxies reside on the SF main sequence, this trend may arise due to a correlation with star-formation rate, and we return to this below.
Similarly, in the right hand side panel of Figure~\ref{fig:tdep_scale}, we show the atomic gas depletion time $t_{\text{dep}}$, as a function of the specific star formation rate (sSFR $\equiv$ SFR/$M_{\ast}$). For this case, we do not find a statistically significant link (Table~\ref{table:galaxy-prop-tdep-stats-delay}).

\begin{figure}
    \centering
	\includegraphics[width=\columnwidth]{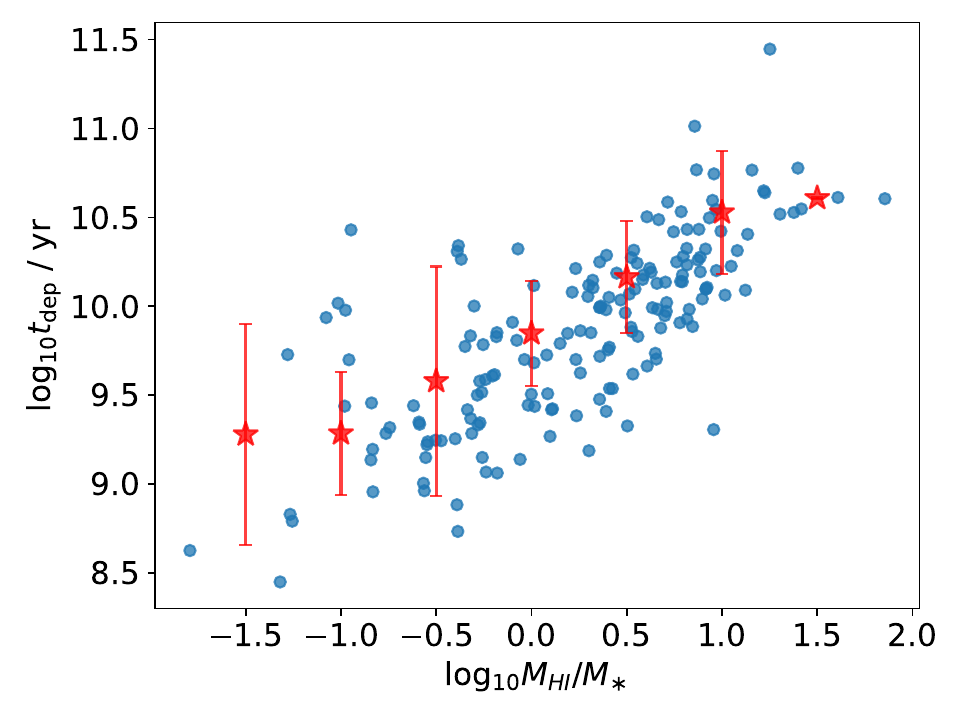}
    \caption{Atomic gas depletion time $t_{\text{dep}}$ as a function of $M_{\text{H{\sc i}}}/M_{\ast}$ for the exponentially-delayed SFH. The red stars represent the running median of the atomic gas depletion $t_{\text{dep}}$ as a function of $M_{\text{H{\sc i}}}/M_{\ast}$ from the MIGHTEE-H{\sc i} data. The error bars on the running median are calculated by using the standard deviation of the value in each bin.}
    \label{fig:mhi-sfr-mhi-ms}
\end{figure}

In Figure \ref{fig:mhi-sfr-mhi-ms}, we show the atomic gas depletion time $t_{\text{dep}}$ with the H{\sc i}-to-stellar mass ratio. As can be seen, galaxies with higher H{\sc i}-to-stellar mass ratios have longer depletion times. This is also confirmed by the correlation tests with both returning  statistically significant values (Table~\ref{table:galaxy-prop-tdep-stats-delay}).

Given the correlations between the atomic gas depletion time and both the stellar mass, the H{\sc i}-to-stellar mass ratio and the SFR, we investigate which is the dominant parameter that controls the gas depletion time. We therefore perform a partial correlation test \citep{Macklin_1982, Johnson_2002, Whittaker_2009} between the H{\sc i} mass, the stellar mass and the SFR of the galaxies. We carry out these tests across all combinations of the three parameters. First, for the partial correlation of SFR vs $M_{\text{H{\sc I}}}$ in the presence of $M_{\star}$ we find the rank coefficient $r = 0.162$, with a $p-$value of $0.033$, thus the correlation between SFR vs $M_{\text{H{\sc I}}}$ is relatively weak, when accounting for stellar mass.
Second, for $M_{\star}$ vs $M_{\text{H{\sc I}}}$ in the presence of SFR,  we find a much more significant partial correlation coefficient $r = 0.54$, with a $p-$value of $10^{-14}$. Finally for the partial correlation of $M_{\star}$ vs SFR in the presence of $M_{\text{H{\sc I}}}$, we find a partial correlation coefficient $r = 0.331$, with a $p-$value of $10^{-6}$. This implies that the stellar mass is the primary driver for gas depletion time (and the SFR).

%In Figure \ref{fig:mhi-sfr-tform}, we compare the atomic gas depletion time, $t_{\text{dep}}$ with the peak time of star formation $t(z_{\text{form}})$ for each galaxy. In this case, no correlation is found, with the points are randomly distributed across different timescales of depletion. The correlation tests confirm this hypothesis (see Table \ref{table:galaxy-prop-tdep-stats-delay}). The peak time of star formation is likely influenced by factors such as galaxy interactions, mergers, and the availability of cold gas in the surrounding environment, and not necessarily just from within the galaxy itself \citep{Leroy_2008, Popping_2014, Somerville_2015, Diaz-Garcia_2020}.

We find a clear result that gas-rich galaxies with higher H{\sc i} mass fractions have longer atomic gas depletion times. This means that they can sustain star formation at their current rate for a longer period of time before their gas is completely depleted. This is consistent with the idea that gas-rich galaxies have a higher potential for ongoing star formation and can replenish their gas reservoirs through accretion from their immediate surroundings \citep[i.e. their CGM;][who also use H{\sc i} selected samples in their studies]{Jaskot_2015, Lutz_2017}.

\begin{figure*}
\begin{subfigure}[b]{\columnwidth}
  \centering
  % include first image
  \captionsetup{justification=centering}
  \includegraphics[width=1\linewidth]{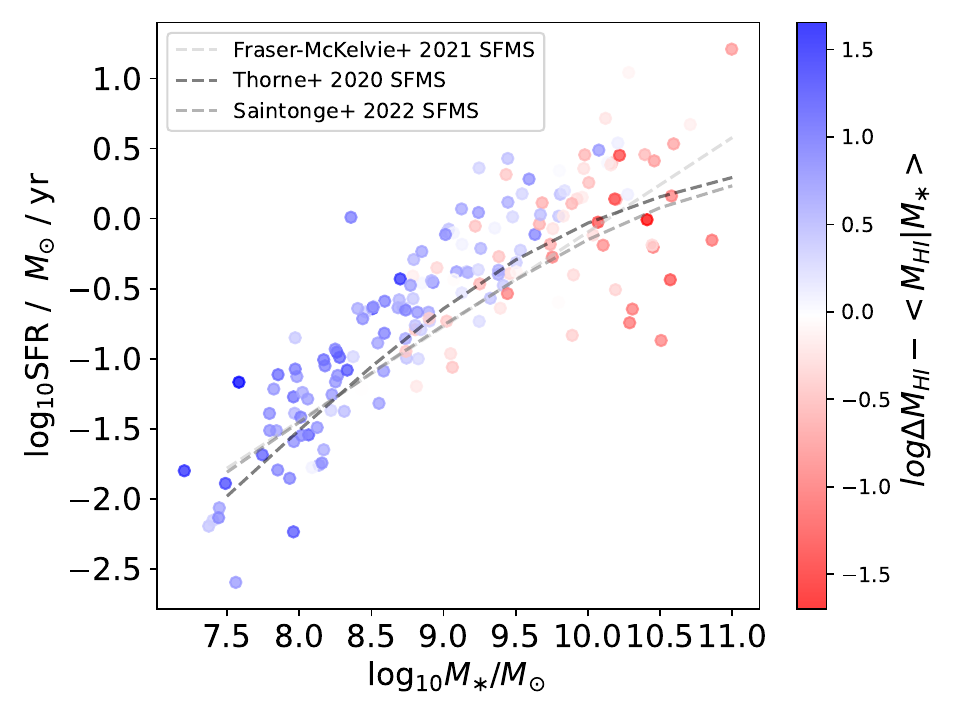} 
  \phantomcaption
  \label{subfig:ms_sfr_mhi_ms}
\end{subfigure}
\begin{subfigure}[b]{\columnwidth}
  \centering
  % include second image
  \captionsetup{justification=centering}
  \includegraphics[width=1\linewidth]{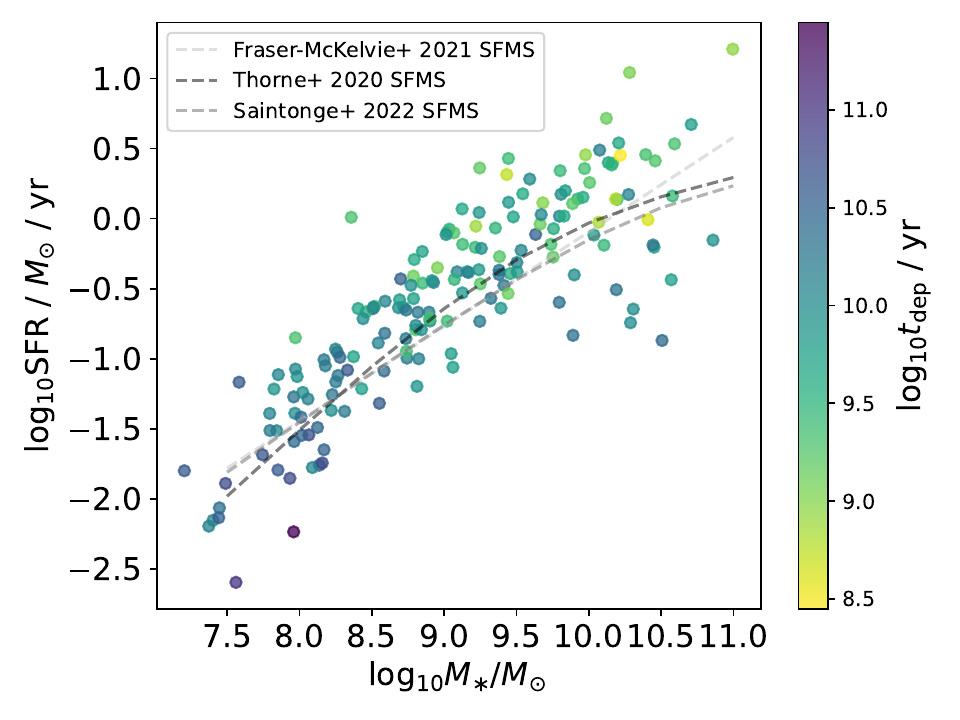}
  \phantomcaption
  \label{subfig:ms_sfr_tdep}
\end{subfigure}
\caption{Variations in $\Delta M_{\text{H{\sc i}}} - < M_{\text{H{\sc i}}} | M_{\ast}>$ (left) and $t_{\text{dep}}$ (right) across the SFR–$M_{\ast}$ plane. The three dashed lines are different SF-MS fits: the black one is the SF-MS as presented in \citet{Thorne_2021}, the light gray one is the linear SF-MS as presented in \citet{Fraser-McKelvie_2021} and the dark grey one is the SF-MS as presented in \citet{Saintonge_2022}.}
  \label{fig:ms_sfr}
\end{figure*}

The observed correlation between stellar mass and atomic gas depletion time also suggests that more massive galaxies not only have higher gas masses but also exhibit a more efficient utilisation of their gas reservoirs for star formation. As dwarf galaxies form in lower potential wells, the gas will experience a lower gravitational pull, allowing more gas to escape due to the stronger effect of stellar feedback in these galaxies \citep{Hopkins_2014, Romano_2023}, 
or be stripped from it, delaying collapse and preventing efficient replenishment on to the galaxy. This culminates in a less efficient conversion rate from H{\sc i} to H$_{2}$ \citep{Hunt_2020}. Other factors, such as the enhanced susceptibility of the dwarf galaxies to the background UV field \citep{Pereira-Wilson_2023}, and their overall lower metallicity \citep{Tremonti_2004} can also reduce the efficiency of the conversion rate. At the other end of the mass scale, the more massive galaxies have a higher density of gas in the central regions that is retained within the deeper potential well, which facilitates star formation \citep{Saintonge_2012, Parkash_2018, Saintonge_2022}.

However, there are some studies which appear to contradict our results. \cite{Schiminovich_2010} find that the inverse of the atomic gas time depletion, star formation efficiency ($\text{SFE} \equiv 1/t_{\text{dep}}$) of massive galaxies in an H{\sc i}-selected sample to be constant at a value of $10^{-9.5}$~yr$^{-1}$ (for a stellar mass range of $10.0 $M$_{\sun} < \log_{10}{M_{\ast}} < 11.5$~M$_{\sun}$). We note however that this is consistent with our results as the limited range in stellar mass in the study of \cite{Schiminovich_2010} means that they would not see the increase (decrease) in gas-depletion time (star-formation efficiency) towards lower stellar masses ($M_{\ast} < 10^{10}$\,M$_{\odot}$).

\cite{Wong_2016} find the SFE to be constant at $10^{-9.65}$~yr$^{-1}$ across five orders of magnitude of stellar mass (range of $7.0$ M$_{\sun} < \log_{10}{M_{\ast}} < 11.5$ M$_{\sun}$) for star-forming galaxies from the SINGG and SUNGG \citep{Meurer_2006} surveys. 
Much of this apparent disagreement can be explained by the differing selection criteria, with the galaxies in these studies being stellar mass selected, rather than selected on their atomic gas mass. 
This is also demonstrated in \citet{Parkash_2018}, who use three different selection criteria for their samples: an H{\sc i}-selected sample, a spiral galaxy selected sample, and a stellar mass selected sample, which shows that H{\sc i}-selected samples will result in measurements of high median H{\sc i} masses and low median SFEs (high median atomic gas depletion times), as opposed stellar mass selected samples. Indeed, \cite{Schiminovich_2010, Huang_2012a} have shown that in the optically selected samples they find on average SFEs which are three times lower than in H{\sc i} selected samples. \

Returning to our measurement of the atomic gas depletion time, using the left hand panel of Figure~\ref{fig:tdep_scale} and restricting our stellar mass range to be closer to the stellar mass ranges in these works ($9.0 $~M$_{\sun} < \log_{10}{M_{\star}} < 11.0 $~M$_{\sun}$), we find that the $t_{\text{dep}}$ trend becomes flatter, with an average SFE of $10^{-9.82 \pm 0.54}$~yr$^{-1}$. 
%This shows that we are biased towards higher gas depletion times, which is a consequence of using an H{\sc i}-selected sample. %. 
Hence, there is a part of the parameter space (i.e. the bottom left corner in the left-hand panel of Figure~\ref{fig:tdep_scale}) where there is a real lack of galaxies (i.e. we could detect highly star-forming galaxies with low stellar mass and low H{\sc i} masses if they existed in abundance). 
This is consistent with these other results, since the H{\sc i}-selected galaxy sample studies \citep{Huang_2014, Jaskot_2015, Lutz_2017, Zhou_2018} find a non-constant, increasing SFE (decreasing $t_{\text{dep}}$) with increasing stellar mass.
On the other hand, stellar-mass or optically-selected samples have a large population of low-stellar mass objects and the scatter in the star-formation main sequence means that many of these lie above the MS and have already used a significant fraction of their neutral gas reservoir \citep{Schiminovich_2010, Wong_2016}.% and find a constant SFE ($t_{\text{dep}}$).

%\begin{figure}
%    \centering
%	\includegraphics[width=\columnwidth]{figures/mhi_sfr_tform_delay.pdf}%
%    \caption{A scatter plot illustrating the atomic gas depletion time $t_{\text{dep}}$ as a function of the peak of star formation $t_{\text{form}}$ for the exponentially-delayed SFH. The dotted blue line represents the running median of the atomic gas depletion $t_{\text{dep}}$ as a function of $t(z_{\text{form}})$. The error bars on the median are calculated by using the $16$th and $84$th percentile values of $t(z_{\text{form}})$ and $t_{\text{dep}}$.}
%    \label{fig:mhi-sfr-tform}
%\end{figure}

\subsubsection{The star-forming main sequence of H{\sc i}-selected galaxies}
Finally, we look at the star-forming main sequence. In Figure~\ref{fig:sfms}, we see that our H{\sc i}-selected galaxies tend to lie above the $z\sim 0$ galaxy main sequence, particularly towards higher mass. 
We also find that the H{\sc i} mass generally increases with stellar mass, as has been found previously with H{\sc i}-selected samples \citep[e.g.][]{Maddox_2015, Pan_2023}.

To investigate this further, in Figure~\ref{fig:ms_sfr} (left) we show the SFR-$M_{\ast}$ plane colour-coded by the deviation from the mean $M_{\ast}-M_{{\text{H\sc i}}}$ relation, using Model B from \citet{Pan_2023} for the mean relation. 
%and by the atomic gas depletion time $t_{\text{dep}}$ (right). 
We find galaxies with a higher positive deviation from the median relation at the lower-mass end of the plot, whilst galaxies tend to have higher negative deviations at the high stellar mass end, which implies that the high-stellar-mass galaxies in our H{\sc i}-selected sample are relatively deficient in H{\sc i} compared to their lower-mass counterparts.
This is consistent with the partial correlation results, where we found that between stellar mass, H{\sc i} mass and SFR, the stellar mass is the main driver of the correlation. Thus, those H{\sc i}-selected galaxies at $M_{\ast} \gtrsim 10^9$M$_{\odot}$, which predominantly lie above the MS have much shorter gas depletion timescales, suggesting that they are using up their gas supply much more efficiently than the galaxies with $M_{\ast} \lesssim 10^9$M$_{\odot}$. 

Similarly, low-mass galaxies tend to have a higher atomic gas depletion time $t_{\text{dep}}$, with depletion time decreasing along the main sequence to high stellar masses. It follows that galaxies with higher stellar masses have likely formed stars earlier in their evolution and have used up a larger portion, but with a significant amount still remaining, of their originally much larger gas reservoirs. This suggests that the most massive galaxies have had a disproportionately large H{\sc i} reservoir in the past in order to form the current mass in stars. Alternatively, they may have been constantly accreting new gas, as the gas depletion time does not account for this effect. As a galaxy releases both energy and material back into its ISM through processes such as stellar winds, supernova explosions and AGN feedback \citep[e.g.][]{Dekel_1986, Kauffmann_2003, Bertone_2007}, gas can be heated and expelled into the galaxy's halo or into the CGM. Over time, the outflowing material can cool and condense, transitioning back into gas that falls back toward the galactic disk \citep{Werk_2014, Tumlinson_2017}. This recycled gas, once reintegrated into the galaxy, can once again form molecular clouds, fuelling new star formation.

However, what is also clear from this figure is that galaxies that begin to drop off from the main sequence at high masses still retain a significant reservoir of H{\sc i}, whilst their star-formation rate is reducing. This suggests a disconnection of the star formation occurring in these galaxies from the availability of neutral atomic hydrogen. Obviously, star-formation arises from the cool molecular gas so the reason for the truncation of star formation relative to a high level of neutral gas could be explained by a reduced efficiency of converting neutral atomic gas to molecular gas, or a reduction in the efficiency of producing stars from the molecular gas reservoir. Several studies have shown that the latter is certainly possible in the highest mass systems, with either significant rotation in the cores of galaxies where the majority of the dense molecular gas resides \citep[e.g.][]{Davis2014}, or where high velocity dispersion prevents that gas from collapsing \citep[e.g.][]{Dey2019}. Such galaxies appear able to retain a significant amount of low-density atomic hydrogen \citep{Serra2012} \cite[see also ][]{Saintonge_2022}.

As H{\sc i}-selected galaxies are often found in less dense environments, such as the outskirts of clusters or in the field (i.e. filaments, walls, sheets), where gas stripping processes are less prevalent \citep{Verdes-Montenegro_2001, Walker_2016, crone-odekon_2018, Jones_2018}. This means that the galaxies in an H{\sc i}-selected sample might be under-represented in dense environments \citep{Basilakos_2007, Meyer_2007, Martin_2012}, i.e. be a biased subset that preferentially retain their gas reservoirs. 
%In fact, it has been shown that H{\sc i} selected samples tend to be the least clustered population \citep{Meyer_2007, Martin_2012}.  

Furthermore, as H{\sc i} extends to larger radii than stars in galaxies, it is more easily perturbed during tidal interactions and hence, more sensitive to external influences that could be caused by the cosmic web \citep{Yun_1994, Chung_2009}. As shown in the section above, we find that the atomic gas depletion time $t_{\text{dep}}$ varies with stellar mass. The timescales it varies across are of order $\sim$~Gyr, as we find an average of $t_{\text{dep}} = 6.9$~Gyr across the whole sample. 
The time for low-redshift galaxies to oscillate across the main sequence to result in the scatter that is observed has been estimated to be around $\sim 5$~Gyr \citep[and decreasing for higher redshifts;][]{Tacchella_2016}. However, our results suggest that the H{\sc i} reservoir could not respond quickly enough to explain the scatter without external influences or additional physical processes occurring in the galaxies that either trigger or delay star formation.
This falls in line with the scenario that the scatter in the main sequence varies on much longer timescales, and it is the environment (i.e. the haloes the galaxies reside in, or their position in the cosmic web) that affects how the galaxies move across the main sequence \citep{Matthee_2019, Berti_2021}. However, the depletion time of a galaxy’s gas reservoir and its consumption time are not the same, as the depletion time only accounts for consumption by star formation. In low-mass star-forming galaxies, most simulations suggest strong outflows that can carry out $10-100$ times more mass than is forming into stars \citep{Nelson_2019, Mitchell_2020, Pandya_2021}. This implies that the gas consumption time would be lower than the depletion time by this factor, making the timescale over which the H{\sc i} reservoir is consumed to be fairly short, and suggests that it may be possible for the lower-mass galaxies to vary around the H{\sc i} main sequence on $\sim$Gyr timescales.

%As well, the H{\sc i}-selected SFR-$M_{\ast}$ plane is also consistent with previous works \citep[e.g.][]{Huang_2014, Saintonge_2017, Saintonge_2022}

\subsection{Environmental effects}
\label{subsec:mergers}

%However, it must be noted that due to our sample selection, as discussed above, it is not a surprising result. We are already biased towards higher gas depletion times as we sample gas-rich systems.

\begin{figure}
    \centering
	\includegraphics[width=1\linewidth]{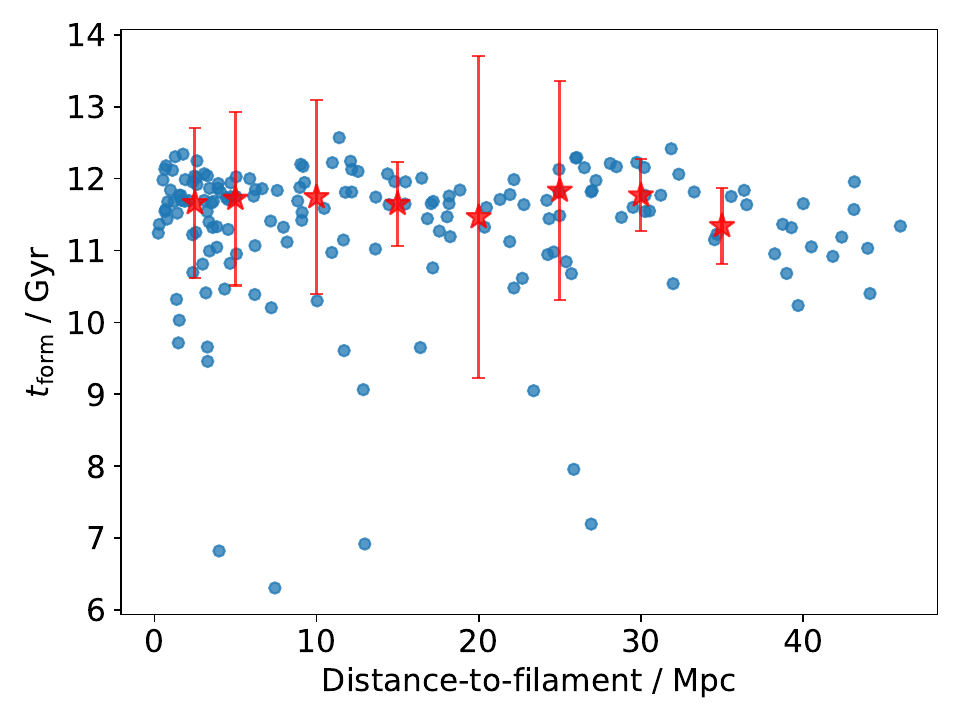} 
    \caption{The time of formation $t_{\text{form}}$ as a function of the distance to the closest filament of a galaxy. The red stars represent the running median of the distance to filament of a galaxy as a function of $t_{\text{form}}$. The error bars on the running median are calculated by using the standard deviation of the value in each bin.}
    \label{fig:dfil-tform}
\end{figure}

As discussed above, the depletion timescale of the H{\sc i} gas cannot be solely responsible for the scatter of the star-formation main sequence, and environmental processes are likely to be important. In Figure~\ref{fig:dfil-tform} we show the formation time of a galaxy and the distance to its nearest filament, using \textsc{Disperse} \citep{sousbie} to determine the skeleton of the cosmic web based on the distribution of galaxies from the COSMOS and XMM-LSS fields, as defined in \cite{tudorache2022}. We find no evidence for any links between the proximity of a galaxy to its nearest filament and when it formed the bulk of its stars (Spearman rank coefficient $r = -0.095$ with $p=0.171$) .

\begin{figure}
    \centering
	\includegraphics[width=\columnwidth]{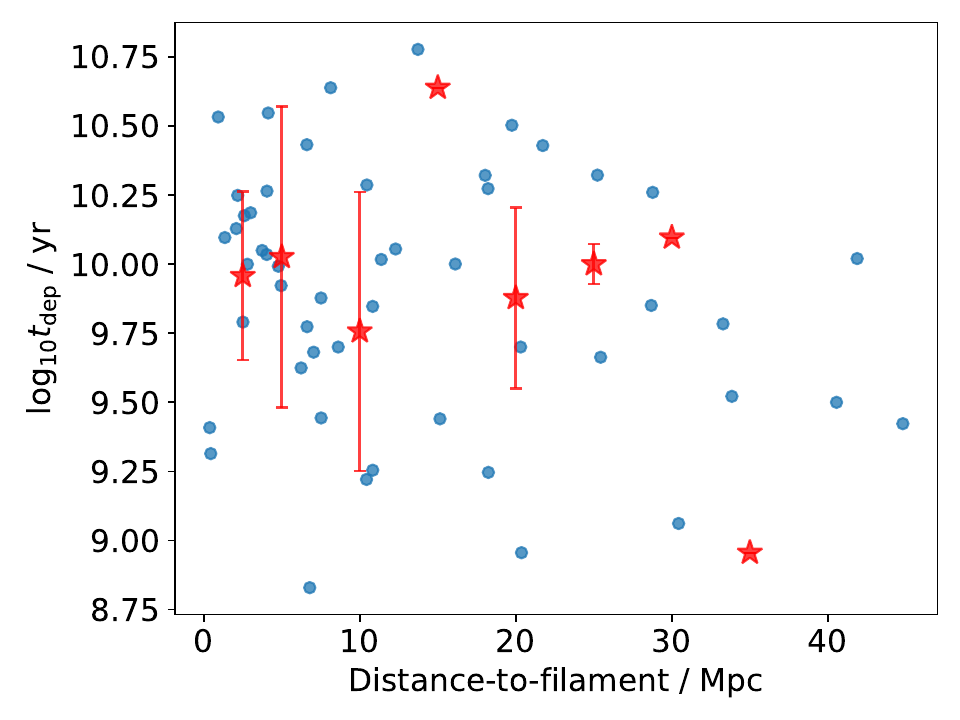}
    \caption{Atomic gas depletion time $t_{\text{dep}}$ as a function of distance to the closest filament of a galaxy. The red stars represent the running median of the distance to the closest filament of a galaxy as a function of $t_{\text{dep}}$. The error bars on the running median are calculated by using the standard deviation of the value in each bin. }
    \label{fig:mhi-sfr-dist}
\end{figure}

Similarly, when investigating a link between the atomic gas depletion time $t_{\text{dep}}$ and the distance-to-filament, we find no correlation - as can be seen in Figure~\ref{fig:mhi-sfr-dist}. 
Therefore, the timescales of star formation in a galaxy do not seem to have any link to its position with respect to the cosmic filaments. 
%Other studies find that older galaxies tend, in fact, to be found more preferentially in clusters as opposed to filaments/voids \citep[e.g.][]{Bernardi_2006, Smith_2012, Chen_2017}, as they had more time to evolve. 
We also find no evidence that the filaments have an effect on how quickly the galaxies convert their neutral gas into stars using the atomic gas depletion time as a proxy for this process, or that the gas is being replenished preferentially in the filamentary environment. 

As discussed in \cite{tudorache2022}, the H{\sc i}-to-stellar mass ratio of a galaxy is related to the filaments of the cosmic web, such that galaxies that have their spins misaligned with their closest filament have higher H{\sc i}-to-stellar mass ratios. As gas-rich mergers are expected to increase the amount of neutral gas in galaxies \citep{Ellison2018}, this could suggest that galaxies which have recently undergone such a merger might have their spin-orientation disrupted with respect to the filament, whereas those galaxies which have not undergone a recent merger will tend to retain their alignment and their evolution is dictated by secular processes.

First of all, we checked if there are any galaxies within the sample showing morphological evidence of undergoing mergers using optical images discussed in Section~\ref{sec:data}. We find only 7 of our galaxies (of which only $4$ are within $10$~Mpc of a filament) show evidence of merger activity \citep{Rajohnson_2022}. They are also randomly distributed in terms of the spin-filament alignment (they do not have a preference for either aligned or misaligned). 

Thus, here we investigate whether galaxies that are aligned exhibit any evidence for a different evolutionary history than their misaligned counterparts. We take the $16$ galaxies from \cite{tudorache2022} that lie within $5$\,Mpc of a filament and investigate if there is any difference in the dot product of the spin vector of the galaxy and the 3-dimensional direction of the filament ($\lvert \cos \phi \rvert$) and the formation time of the H{\sc i}-selected galaxies (see \citealt{tudorache2022} for how the angle is calculated).  

\begin{figure}
    \centering
	\includegraphics[width=\columnwidth]{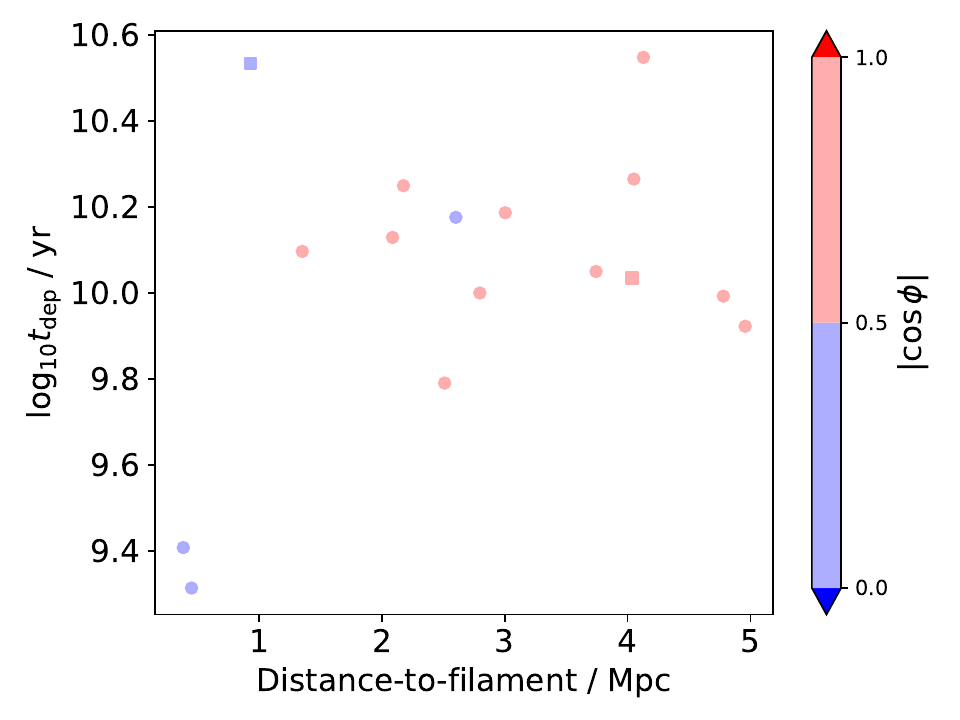}
    \caption{Atomic gas depletion time, $t_{\text{dep}}$ as a function of its distance to the closest filament of a galaxy, colour-coded by spin-filament cos-angle $\lvert \cos \phi \rvert$. The two squares represent the galaxies which presented hints of mergers by visually inspecting the optical images.}
    \label{fig:spin-tdep}
\end{figure}

Figure~\ref{fig:spin-tdep} shows that the two H{\sc i} galaxies that reside closest to the spine of the filament, which also have their spin axes misaligned with the filament direction, also have the lowest gas-depletion timescales. A 2-D Kolmogorov-Smirnov \citep{Peacock1983} shows that the misaligned ($\lvert \cos \phi \rvert < 0.5$) and aligned ($\lvert \cos \phi \rvert > 0.5$) galaxy sub-samples in the $t_{\rm dep}$ -- distance to filament plane have a p-value of $0.002$, suggesting that the difference is statistically significant, even with such low numbers.
However, given the very small number of galaxies, at this point we suggest this may be an indication that galaxies that are towards the spines of filaments may have the spins axes disrupted coupled with an increase in the star-formation rate. A gas-rich merger could obviously explain both such observations, but confirmation of this result will require a much larger sample size.

\begin{table}
\centering
\caption{The coefficients and p-values for the two correlation tests, Kendall's Tau and Spearman Rank, for each parameter against the $t_{\text{form}}$ for the delayed SFH.}
\label{table:galaxy-prop-stats-delay}
\begin{tabular}{ccccc}
\hline\hline
 \multirow{2}{*}{\textbf{Parameter}  }                             & \multicolumn{2}{c}{\textbf{Kendall's Tau}} & \multicolumn{2}{c}{\textbf{Spearman Rank}} \\ 
& \textbf{$\tau$}   & \textbf{p-value}  & \textbf{coefficient}   & \textbf{p-value}  \\ \hline\hline
Distance & $-0.054$ & $0.287$ & $-0.082$ & $0.279$\\
$\lvert \cos \phi \rvert$ & $0.023$ & $0.814$ & $0.023$ & $0.872$\\
$M_{\mathrm{HI}}$ & $-0.112$ & $0.028$ & $-0.17$ & $0.024$\\
$M_{\ast}$ & $-0.313$ & $0.0$ & $-0.458$ & $0.0$\\
$M_{\mathrm{HI}}/M_{\ast}$ & $0.304$ & $0.0$ & $0.444$ & $0.0$\\
\hline
\end{tabular}
\end{table}

\begin{table}
\centering
\caption{The coefficients and p-values for the two correlation tests, Kendall's Tau and Spearman Rank, for each parameter against the $t_{\text{dep}}$ for the delayed SFH.}
\label{table:galaxy-prop-tdep-stats-delay}
\begin{tabular}{ccccc}
\hline\hline
 \multirow{2}{*}{\textbf{Parameter}}                             & \multicolumn{2}{c}{\textbf{Kendall's Tau}} & \multicolumn{2}{c}{\textbf{Spearman Rank}} \\ 
& \textbf{$\tau$}   & \textbf{p-value}  & \textbf{coefficient}   & \textbf{p-value}  \\ \hline\hline
Distance & $0.175$ & $0.001$ & $0.257$ & $0.001$\\
$\lvert \cos \phi \rvert$ & $-0.054$ & $0.575$ & $-0.06$ & $0.676$\\
$M_{\mathrm{HI}}$ & $0.02$ & $0.698$ & $0.033$ & $0.661$\\
$M_{\ast}$ & $-0.368$ & $0.0$ & $-0.522$ & $0.0$\\
$M_{\mathrm{HI}}/M_{\ast}$ & $0.541$ & $0.0$ & $0.713$ & $0.0$\\
SFR & $-0.503$ & $0.0$ & $-0.692$ & $0.0$\\
sSFR & $-0.018$ & $0.717$ & $-0.022$ & $0.774$\\
$t_{\text{form}}$ & $-0.018$ & $0.719$ & $-0.025$ & $0.738$\\
\hline
\end{tabular}
\end{table}

\section{Conclusions}
\label{sec:conclusions}

We have fitted the SEDs of $187$ galaxies from the MIGHTEE-H{\sc i} Early Release data with spectroscopic redshifts and excellent multi-wavelength photometric data from the FUV to the FIR in order to investigate links between the H{\sc i} content of galaxies, the large-scale structures of the cosmic web and star formation. By using \textsc{Bagpipes}, a Bayesian inference-based SED fitting code, we find that most of the galaxies are best described by either a delayed-exponentially declining, a lognormal or an exponentially declining model for the SFHs.
We then employed the SED modelling, utilising the unprecedented wealth of multi-wavelength data for a H{\sc i}-selected samples, to measure the stellar mass, SFR, and the peak time of star formation $t_{\text{form}}$, to interpret the process of star formation and the effect of the environment in the context of an H{\sc i}-selected sample.

\subsection{H{\sc i} galaxy properties}
In terms of the H{\sc i} properties with respect to the SFHs, our main findings are:
\begin{itemize}
    \item We find correlations between the H{\sc i}-to-stellar mass ratio and the time of formation (positive correlation) and stellar mass and the time of formation (anti-correlation);
    \item We find positive correlations between the atomic gas depletion time and H{\sc i}-to-stellar mass ratio and stellar mass of a galaxy;
    \item We find no correlations between the atomic gas depletion time and the sSFR of a galaxy
\end{itemize}

These results are consistent with the picture that lower mass, gas-rich galaxies have a higher depletion time due to a shallower potential well and less efficient star formation, whilst more massive galaxies have already depleted their gas and formed stars efficiently. This is also reflected in the partial correlation findings, from which we infer that the stellar mass is the main driver. This picture is often seen in H{\sc i} selected samples, so care must be taken when drawing this conclusion. Most importantly, this persists independent of the SFH model used to compute the quantities involved, suggesting the pathway that a galaxy takes to form its stars does not necessarily affect the final correlations. Moreover, due to the timescales of the atomic gas depletion time, we conclude that the scatter in the star forming main sequence cannot be caused by short timescale variations, and it has to be caused by long term effects, such as the ones that could be caused by the environment, although we cannot rule out continual short-term expulsion and reintegration of the gas reservoir, particularly in the lower-mass systems.

\subsection{Links to the filamenary environment}

Investigating the link between the galaxy properties and the large-scale environment, especially the spin-filament alignment and SFHs, the main findings are:
\begin{itemize}
    \item By visually inspecting the SFHs of a $7$ galaxy sub-sample with visible hints of on-going mergers, we do not find any signs of recent episodes of star formation in those that misaligned with their nearest filament; 
    \item For the $51$ galaxies sub-sample where we could calculate the spin-filament alignment cos-angle, $\lvert \cos \phi \rvert$, we find that distance-to-filament does not correlate with the atomic gas depletion time $t_{\text{dep}}$;
    \item Interestingly, we find the two galaxies that lie the cloest to their filament ($< 1$\,Mpc) have the shortest H{\sc i} depletion times, which may be indicative of a link between the galaxy properties and its filamentary environment.
    % Other than the two H{\sc i} galaxies which reside closest to the filament, we find no significant trend between the atomic gas depletion time $t_{\text{dep}}$ and the spin-filament alignment cos-angle, $\lvert \cos \phi \rvert$ for the galaxies closest to a cosmic filament.
\end{itemize}

These conclusions could be explained in two ways: 
either mergers are not the cause for the split in the spin-filament alignment, such that the H{\sc i}-to-stellar mass ratio is a secondary correlation to another parameter, such as age, or; the photometric data alone cannot determine the SFHs accurately enough such that a merger history would be observed.
%However, the hypothesis that we can use SFHs to track explain the difference in the spin-filament alignment as a function of H{\sc i}-to-stellar mass ratio cannot be excluded. This is due to the fact that we only use photometric data to infer our SFHs, and they are dependent on the models we chose. 
To investigate this question further, larger samples or more precise spectroscopic data would be required. We will soon have access to far more data from the range of current H{\sc i} surveys currently underway, but in particular, the wealth of ancillary data over the MIGHTEE fields and the Looking at the Distant University with the MeerKAT Array \citep[LADUMA; ][]{Blyth2016} Survey will help elucidate these questions further.
Spectroscopy would  also provide a different avenue into better constrained SFHs. As such, we could examine the SFHs of galaxies and look for signs of ongoing star formation or merger events by using more accurate measurements of the stellar populations \citep{Gallazzi_2009, Cappellari_2023, Nersesian_2024}. The forthcoming Wide-area VISTA Extragalactic Surveys \citep[WAVES; ][]{WAVES} and specifically the Optical, Radio Continuum and HI Deep Spectroscopic Survey \citep[ORCHIDSS; ][]{ORCHIDSS} will greatly enhance our ability to extract key information for these H{\sc i} rich galaxies, while facilitating stacking approaches for stellar-mass or SFR selected samples.
%By combining data from different sources and using an IFU, we would be able track the SFHs in these galaxies and confirm if past mergers are indeed responsible for the misalignment between their hydrogen gas disks and the cosmic filaments. This would enable us to determine whether there is any evidence for mergers on a galaxy-by-galaxy basis due to the significant improvement in the quality of the spectral data.}

\section*{Acknowledgements}
MNT, MJJ, IH, SLJ, and HP acknowledge the support of a UKRI Frontiers Research Grant [EP/X026639/1], which was selected by the European Research Council.
MNT and MJJ acknowledge support from the Oxford Hintze Centre for Astrophysical Surveys which is funded through generous support from the Hintze Family Charitable Foundation. IH, MJJ and AAP acknowledge support from the UK Science and Technology Facilities Council [ST/N000919/1]. 
IH acknowledges support from the South African Radio Astronomy Observatory which is a facility of the National Research Foundation (NRF), an agency of the Department of Science and Innovation. The MeerKAT telescope is operated by the South African Radio Astronomy Observatory, which is a facility of the National Research Foundation, an agency of the Department of Science and Innovation. We acknowledge use of the Inter-University Institute for Data Intensive Astronomy (IDIA) data intensive research cloud for data processing. IDIA is a South African university partnership involving the University of Cape Town, the University of Pretoria and the University of the Western Cape. The authors acknowledge the Centre for High Performance Computing (CHPC), South Africa, for providing computational resources to this research project. This work is based on data products from observations made with ESO Telescopes at the La Silla Paranal Observatory under ESO programme ID 179.A-2005 (Ultra-VISTA) and ID 179.A- 2006 (VIDEO) and on data products produced by CALET and the Cambridge Astronomy Survey Unit on behalf of the Ultra-VISTA and VIDEO consortia. Based on observations obtained with MegaPrime/MegaCam, a joint project of CFHT and CEA/IRFU, at the Canada-France-Hawaii Telescope (CFHT) which is operated by the National Research Council (NRC) of Canada, the Institut National des Science de l’Univers of the Centre National de la Recherche Scientifique (CNRS) of France, and the University of Hawaii. This work is based in part on data products produced at Terapix available at the Canadian Astronomy Data Centre as part of the Canada-France-Hawaii Telescope Legacy Survey, a collaborative project of NRC and CNRS. The Hyper Suprime-Cam (HSC) collaboration includes the astronomical communities of Japan and Taiwan, and Princeton University. The HSC instrumentation and software were developed by the National Astronomical Observatory of Japan (NAOJ), the Kavli Institute for the Physics and Mathematics of the Universe (Kavli IPMU), the University of Tokyo, the High Energy Accelerator Research Organization (KEK), the Academia Sinica Institute for Astronomy and Astrophysics in Taiwan (ASIAA), and Princeton University. Funding was contributed by the FIRST program from Japanese Cabinet Office, the Ministry of Education, Culture, Sports, Science and Technology (MEXT), the Japan Society for the Promotion of Science (JSPS), Japan Science and Technology Agency (JST), the Toray Science Foundation, NAOJ, Kavli IPMU, KEK, ASIAA, and Princeton University. 

This research made use of Astropy,\footnote{\url{http://www.astropy.org}} a community-developed core Python package for Astronomy \citep{astropy:2013, astropy:2018}.

%%%%%%%%%%%%%%%%%%%%%%%%%%%%%%%%%%%%%%%%%%%%%%%%%%
\section*{Data Availability}

The MIGHTEE-H{\sc i} spectral cubes will be released as part of the first data release of the MIGHTEE survey, which include cubelets of the sources discussed in this paper \citep{Heywood_2022}. The derived quantities from the multi-wavelength ancillary data were released with the final data release of the VIDEO survey mid 2021. Alternative products are already available from the Deep Extragalactic VIsible Legacy Survey \citep[DEVILS;][]{devils}.



\bibliographystyle{mnras}
\bibliography{ref}

\begin{thebibliography}{}
\makeatletter
\relax
\def\mn@urlcharsother{\let\do\@makeother \do\$\do\&\do\#\do\^\do\_\do\%\do\~}
\def\mn@doi{\begingroup\mn@urlcharsother \@ifnextchar [ {\mn@doi@} {\mn@doi@[]}}
\def\mn@doi@[#1]#2{\def\@tempa{#1}\ifx\@tempa\@empty \href {http://dx.doi.org/#2} {doi:#2}\else \href {http://dx.doi.org/#2} {#1}\fi \endgroup}
\def\mn@eprint#1#2{\mn@eprint@#1:#2::\@nil}
\def\mn@eprint@arXiv#1{\href {http://arxiv.org/abs/#1} {{\tt arXiv:#1}}}
\def\mn@eprint@dblp#1{\href {http://dblp.uni-trier.de/rec/bibtex/#1.xml} {dblp:#1}}
\def\mn@eprint@#1:#2:#3:#4\@nil{\def\@tempa {#1}\def\@tempb {#2}\def\@tempc {#3}\ifx \@tempc \@empty \let \@tempc \@tempb \let \@tempb \@tempa \fi \ifx \@tempb \@empty \def\@tempb {arXiv}\fi \@ifundefined {mn@eprint@\@tempb}{\@tempb:\@tempc}{\expandafter \expandafter \csname mn@eprint@\@tempb\endcsname \expandafter{\@tempc}}}

\bibitem[\protect\citeauthoryear{{Abramson}, {Gladders}, {Dressler}, {Oemler}, {Poggianti}  \& {Vulcani}}{{Abramson} et~al.}{2015}]{Abramson_2015}
{Abramson} L.~E.,  {Gladders} M.~D.,  {Dressler} A.,  {Oemler} Augustus J.,  {Poggianti} B.,   {Vulcani} B.,  2015, \mn@doi [\apjl] {10.1088/2041-8205/801/1/L12}, \href {https://ui.adsabs.harvard.edu/abs/2015ApJ...801L..12A} {801, L12}

\bibitem[\protect\citeauthoryear{Adams, Bowler, Jarvis, Häußler  \& Lagos}{Adams et~al.}{2021}]{adams_2021}
Adams N.~J.,  Bowler R. A.~A.,  Jarvis M.~J.,  Häußler B.,   Lagos C. D.~P.,  2021, \mn@doi [Monthly Notices of the Royal Astronomical Society] {10.1093/mnras/stab1956}, 506, 4933–4951

\bibitem[\protect\citeauthoryear{{Aihara} et~al.,}{{Aihara} et~al.}{2019}]{Aihara2019}
{Aihara} H.,  et~al., 2019, \mn@doi [\pasj] {10.1093/pasj/psz103}, \href {https://ui.adsabs.harvard.edu/abs/2019PASJ...71..114A} {71, 114}

\bibitem[\protect\citeauthoryear{{Astropy Collaboration} et~al.,}{{Astropy Collaboration} et~al.}{2013}]{astropy:2013}
{Astropy Collaboration} et~al., 2013, \mn@doi [\aap] {10.1051/0004-6361/201322068}, \href {http://adsabs.harvard.edu/abs/2013A%26A...558A..33A} {558, A33}

\bibitem[\protect\citeauthoryear{{Astropy Collaboration} et~al.,}{{Astropy Collaboration} et~al.}{2018}]{astropy:2018}
{Astropy Collaboration} et~al., 2018, \mn@doi [\aj] {10.3847/1538-3881/aabc4f}, \href {https://ui.adsabs.harvard.edu/abs/2018AJ....156..123A} {156, 123}

\bibitem[\protect\citeauthoryear{{Baldry}, {Glazebrook}, {Brinkmann}, {Ivezi{\'c}}, {Lupton}, {Nichol}  \& {Szalay}}{{Baldry} et~al.}{2004}]{Baldry_2004}
{Baldry} I.~K.,  {Glazebrook} K.,  {Brinkmann} J.,  {Ivezi{\'c}} {\v{Z}}.,  {Lupton} R.~H.,  {Nichol} R.~C.,   {Szalay} A.~S.,  2004, \mn@doi [\apj] {10.1086/380092}, \href {https://ui.adsabs.harvard.edu/abs/2004ApJ...600..681B} {600, 681}

\bibitem[\protect\citeauthoryear{{Balogh}, {Baldry}, {Nichol}, {Miller}, {Bower}  \& {Glazebrook}}{{Balogh} et~al.}{2004}]{Balogh_2004b}
{Balogh} M.~L.,  {Baldry} I.~K.,  {Nichol} R.,  {Miller} C.,  {Bower} R.,   {Glazebrook} K.,  2004, \mn@doi [\apjl] {10.1086/426079}, \href {https://ui.adsabs.harvard.edu/abs/2004ApJ...615L.101B} {615, L101}

\bibitem[\protect\citeauthoryear{{Barnes} \& {Hernquist}}{{Barnes} \& {Hernquist}}{1991}]{Barnes_1991}
{Barnes} J.~E.,  {Hernquist} L.~E.,  1991, \mn@doi [The Astrophysical Journal, Letters] {10.1086/185978}, \href {https://ui.adsabs.harvard.edu/abs/1991ApJ...370L..65B} {370, L65}

\bibitem[\protect\citeauthoryear{{Basilakos}, {Plionis}, {Kova{\v{c}}}  \& {Voglis}}{{Basilakos} et~al.}{2007}]{Basilakos_2007}
{Basilakos} S.,  {Plionis} M.,  {Kova{\v{c}}} K.,   {Voglis} N.,  2007, \mn@doi [\mnras] {10.1111/j.1365-2966.2007.11781.x}, \href {https://ui.adsabs.harvard.edu/abs/2007MNRAS.378..301B} {378, 301}

\bibitem[\protect\citeauthoryear{{Behroozi}, {Wechsler}  \& {Conroy}}{{Behroozi} et~al.}{2013}]{Behroozi_2013}
{Behroozi} P.~S.,  {Wechsler} R.~H.,   {Conroy} C.,  2013, \mn@doi [\apj] {10.1088/0004-637X/770/1/57}, \href {https://ui.adsabs.harvard.edu/abs/2013ApJ...770...57B} {770, 57}

\bibitem[\protect\citeauthoryear{{Berti}, {Coil}, {Hearin}  \& {Behroozi}}{{Berti} et~al.}{2021}]{Berti_2021}
{Berti} A.~M.,  {Coil} A.~L.,  {Hearin} A.~P.,   {Behroozi} P.~S.,  2021, \mn@doi [\aj] {10.3847/1538-3881/abcc6a}, \href {https://ui.adsabs.harvard.edu/abs/2021AJ....161...49B} {161, 49}

\bibitem[\protect\citeauthoryear{{Bertone}, {De Lucia}  \& {Thomas}}{{Bertone} et~al.}{2007}]{Bertone_2007}
{Bertone} S.,  {De Lucia} G.,   {Thomas} P.~A.,  2007, \mn@doi [Monthly Notices of the Royal Astronomical Society] {10.1111/j.1365-2966.2007.11997.x}, \href {https://ui.adsabs.harvard.edu/abs/2007MNRAS.379.1143B} {379, 1143}

\bibitem[\protect\citeauthoryear{{Blue Bird} et~al.,}{{Blue Bird} et~al.}{2020}]{bluebird_2019}
{Blue Bird} J.,  et~al., 2020, Monthly Notices of the Royal Astronomical Society, 492, 153

\bibitem[\protect\citeauthoryear{{Blyth} et~al.,}{{Blyth} et~al.}{2016}]{Blyth2016}
{Blyth} S.,  et~al., 2016, in MeerKAT Science: On the Pathway to the SKA. p.~4

\bibitem[\protect\citeauthoryear{{Bolzonella}, {Miralles}  \& {Pell{\'o}}}{{Bolzonella} et~al.}{2000}]{Bolzonella_2000}
{Bolzonella} M.,  {Miralles} J.~M.,   {Pell{\'o}} R.,  2000, \mn@doi [\aap] {10.48550/arXiv.astro-ph/0003380}, \href {https://ui.adsabs.harvard.edu/abs/2000A&A...363..476B} {363, 476}

\bibitem[\protect\citeauthoryear{{Bond}, {Kofman}  \& {Pogosyan}}{{Bond} et~al.}{1996}]{bond_1996}
{Bond} J.~R.,  {Kofman} L.,   {Pogosyan} D.,  1996, \mn@doi [Nature Astrophysics] {10.1038/380603a0}, \href {https://ui.adsabs.harvard.edu/abs/1996Natur.380..603B} {380, 603}

\bibitem[\protect\citeauthoryear{{Bruzual} \& {Charlot}}{{Bruzual} \& {Charlot}}{2003}]{bruzual2003}
{Bruzual} G.,  {Charlot} S.,  2003, \mn@doi [\mnras] {10.1046/j.1365-8711.2003.06897.x}, \href {https://ui.adsabs.harvard.edu/abs/2003MNRAS.344.1000B} {344, 1000}

\bibitem[\protect\citeauthoryear{Calzetti, Armus, Bohlin, Kinney, Koornneef  \& Storchi-Bergmann}{Calzetti et~al.}{2000}]{Calzetti2000}
Calzetti D.,  Armus L.,  Bohlin R.~C.,  Kinney A.~L.,  Koornneef J.,   Storchi-Bergmann T.,  2000, \mn@doi [The Astrophysical Journal] {10.1086/308692}, 533, 682

\bibitem[\protect\citeauthoryear{{Capak} et~al.,}{{Capak} et~al.}{2007}]{Capak2007}
{Capak} P.,  et~al., 2007, \mn@doi [\apjs] {10.1086/519081}, \href {https://ui.adsabs.harvard.edu/abs/2007ApJS..172...99C} {172, 99}

\bibitem[\protect\citeauthoryear{{Cappellari}}{{Cappellari}}{2023}]{Cappellari_2023}
{Cappellari} M.,  2023, \mn@doi [Monthly Notices of the Royal Astronomical Society] {10.1093/mnras/stad2597}, \href {https://ui.adsabs.harvard.edu/abs/2023MNRAS.526.3273C} {526, 3273}

\bibitem[\protect\citeauthoryear{{Cardelli}, {Clayton}  \& {Mathis}}{{Cardelli} et~al.}{1989}]{Cardelli_1989}
{Cardelli} J.~A.,  {Clayton} G.~C.,   {Mathis} J.~S.,  1989, \mn@doi [\apj] {10.1086/167900}, \href {https://ui.adsabs.harvard.edu/abs/1989ApJ...345..245C} {345, 245}

\bibitem[\protect\citeauthoryear{{Carnall}, {McLure}, {Dunlop}  \& {Dav{\'e}}}{{Carnall} et~al.}{2018}]{Carnall_2018}
{Carnall} A.~C.,  {McLure} R.~J.,  {Dunlop} J.~S.,   {Dav{\'e}} R.,  2018, \mn@doi [\mnras] {10.1093/mnras/sty2169}, \href {https://ui.adsabs.harvard.edu/abs/2018MNRAS.480.4379C} {480, 4379}

\bibitem[\protect\citeauthoryear{{Cenci}, {Feldmann}, {Gensior}, {Moreno}, {Bassini}  \& {Bernardini}}{{Cenci} et~al.}{2024}]{Cenci_2024}
{Cenci} E.,  {Feldmann} R.,  {Gensior} J.,  {Moreno} J.,  {Bassini} L.,   {Bernardini} M.,  2024, \mn@doi [\mnras] {10.1093/mnras/stad3709}, \href {https://ui.adsabs.harvard.edu/abs/2024MNRAS.527.7871C} {527, 7871}

\bibitem[\protect\citeauthoryear{{Chabrier}}{{Chabrier}}{2003}]{Chabrier_2003}
{Chabrier} G.,  2003, \mn@doi [\pasp] {10.1086/376392}, \href {https://ui.adsabs.harvard.edu/abs/2003PASP..115..763C} {115, 763}

\bibitem[\protect\citeauthoryear{{Chou}, {Bridge}  \& {Abraham}}{{Chou} et~al.}{2013}]{Chou_2013}
{Chou} R.~C.~Y.,  {Bridge} C.~R.,   {Abraham} R.~G.,  2013, in {Sun} W.~H.,  {Xu} C.~K.,  {Scoville} N.~Z.,   {Sanders} D.~B.,  eds,  Astronomical Society of the Pacific Conference Series Vol. 477, Galaxy Mergers in an Evolving Universe. p.~145

\bibitem[\protect\citeauthoryear{{Chung}, {van Gorkom}, {Kenney}, {Crowl}  \& {Vollmer}}{{Chung} et~al.}{2009}]{Chung_2009}
{Chung} A.,  {van Gorkom} J.~H.,  {Kenney} J. D.~P.,  {Crowl} H.,   {Vollmer} B.,  2009, \mn@doi [\aj] {10.1088/0004-6256/138/6/1741}, \href {https://ui.adsabs.harvard.edu/abs/2009AJ....138.1741C} {138, 1741}

\bibitem[\protect\citeauthoryear{Collier, Frank, Sekhar  \& Taylor}{Collier et~al.}{2021}]{Collier_2021}
Collier J.~D.,  Frank B.,  Sekhar S.,   Taylor A.~R.,  2021, in 2021 XXXIVth General Assembly and Scientific Symposium of the International Union of Radio Science (URSI GASS). pp~1--4, \mn@doi{10.23919/URSIGASS51995.2021.9560276}

\bibitem[\protect\citeauthoryear{{Cowie}, {Songaila}, {Hu}  \& {Cohen}}{{Cowie} et~al.}{1996}]{Cowie_1996}
{Cowie} L.~L.,  {Songaila} A.,  {Hu} E.~M.,   {Cohen} J.~G.,  1996, \mn@doi [\aj] {10.1086/118058}, \href {https://ui.adsabs.harvard.edu/abs/1996AJ....112..839C} {112, 839}

\bibitem[\protect\citeauthoryear{{Crone Odekon}, {Hallenbeck}, {Haynes}, {Koopmann}, {Phi}  \& {Wolfe}}{{Crone Odekon} et~al.}{2018}]{crone-odekon_2018}
{Crone Odekon} M.,  {Hallenbeck} G.,  {Haynes} M.~P.,  {Koopmann} R.~A.,  {Phi} A.,   {Wolfe} P.-F.,  2018, \mn@doi [\apj] {10.3847/1538-4357/aaa1e8}, \href {https://ui.adsabs.harvard.edu/abs/2018ApJ...852..142C} {852, 142}

\bibitem[\protect\citeauthoryear{{Croton} et~al.,}{{Croton} et~al.}{2006}]{Croton_2006}
{Croton} D.~J.,  et~al., 2006, \mn@doi [Monthly Notices of the Royal Astronomical Society] {10.1111/j.1365-2966.2005.09675.x}, \href {https://ui.adsabs.harvard.edu/abs/2006MNRAS.365...11C} {365, 11}

\bibitem[\protect\citeauthoryear{{Dav{\'e}}, {Angl{\'e}s-Alc{\'a}zar}, {Narayanan}, {Li}, {Rafieferantsoa}  \& {Appleby}}{{Dav{\'e}} et~al.}{2019}]{simba-sim}
{Dav{\'e}} R.,  {Angl{\'e}s-Alc{\'a}zar} D.,  {Narayanan} D.,  {Li} Q.,  {Rafieferantsoa} M.~H.,   {Appleby} S.,  2019, \mn@doi [\mnras] {10.1093/mnras/stz937}, \href {https://ui.adsabs.harvard.edu/abs/2019MNRAS.486.2827D} {486, 2827}

\bibitem[\protect\citeauthoryear{{Davies} et~al.,}{{Davies} et~al.}{2018}]{devils}
{Davies} L.~J.~M.,  et~al., 2018, \mn@doi [\mnras] {10.1093/mnras/sty1553}, \href {https://ui.adsabs.harvard.edu/abs/2018MNRAS.480..768D} {480, 768}

\bibitem[\protect\citeauthoryear{{Davies} et~al.,}{{Davies} et~al.}{2019}]{Davies_2019}
{Davies} L.~J.~M.,  et~al., 2019, \mn@doi [\mnras] {10.1093/mnras/sty3393}, \href {https://ui.adsabs.harvard.edu/abs/2019MNRAS.483.5444D} {483, 5444}

\bibitem[\protect\citeauthoryear{{Davies} et~al.,}{{Davies} et~al.}{2021}]{devils-photom}
{Davies} L.~J.~M.,  et~al., 2021, \mn@doi [\mnras] {10.1093/mnras/stab1601}, \href {https://ui.adsabs.harvard.edu/abs/2021MNRAS.506..256D} {506, 256}

\bibitem[\protect\citeauthoryear{{Davis}}{{Davis}}{2014}]{Davis2014}
{Davis} T.~A.,  2014, \mn@doi [\mnras] {10.1093/mnras/stu1850}, \href {https://ui.adsabs.harvard.edu/abs/2014MNRAS.445.2378D} {445, 2378}

\bibitem[\protect\citeauthoryear{{Davis} \& {Geller}}{{Davis} \& {Geller}}{1976}]{Davis_Geller_1976}
{Davis} M.,  {Geller} M.~J.,  1976, \mn@doi [\apj] {10.1086/154575}, \href {https://ui.adsabs.harvard.edu/abs/1976ApJ...208...13D} {208, 13}

\bibitem[\protect\citeauthoryear{{De Lucia}, {Springel}, {White}, {Croton}  \& {Kauffmann}}{{De Lucia} et~al.}{2006}]{DeLucia_2006}
{De Lucia} G.,  {Springel} V.,  {White} S. D.~M.,  {Croton} D.,   {Kauffmann} G.,  2006, \mn@doi [\mnras] {10.1111/j.1365-2966.2005.09879.x}, \href {https://ui.adsabs.harvard.edu/abs/2006MNRAS.366..499D} {366, 499}

\bibitem[\protect\citeauthoryear{{Dekel} \& {Mandelker}}{{Dekel} \& {Mandelker}}{2014}]{Dekel_2014}
{Dekel} A.,  {Mandelker} N.,  2014, \mn@doi [\mnras] {10.1093/mnras/stu1427}, \href {https://ui.adsabs.harvard.edu/abs/2014MNRAS.444.2071D} {444, 2071}

\bibitem[\protect\citeauthoryear{{Dekel} \& {Silk}}{{Dekel} \& {Silk}}{1986}]{Dekel_1986}
{Dekel} A.,  {Silk} J.,  1986, \mn@doi [The Astrophysical Journal] {10.1086/164050}, \href {https://ui.adsabs.harvard.edu/abs/1986ApJ...303...39D} {303, 39}

\bibitem[\protect\citeauthoryear{{Dey} et~al.,}{{Dey} et~al.}{2019}]{Dey2019}
{Dey} B.,  et~al., 2019, \mn@doi [\mnras] {10.1093/mnras/stz1777}, \href {https://ui.adsabs.harvard.edu/abs/2019MNRAS.488.1926D} {488, 1926}

\bibitem[\protect\citeauthoryear{{Di Matteo}, {Springel}  \& {Hernquist}}{{Di Matteo} et~al.}{2005}]{DiMatteo_2005}
{Di Matteo} T.,  {Springel} V.,   {Hernquist} L.,  2005, \mn@doi [Nature] {10.1038/nature03335}, \href {https://ui.adsabs.harvard.edu/abs/2005Natur.433..604D} {433, 604}

\bibitem[\protect\citeauthoryear{{Dodson} et~al.,}{{Dodson} et~al.}{2022}]{Dodson_2022}
{Dodson} R.,  et~al., 2022, \mn@doi [\aj] {10.3847/1538-3881/ac3e65}, \href {https://ui.adsabs.harvard.edu/abs/2022AJ....163...59D} {163, 59}

\bibitem[\protect\citeauthoryear{{Dressler}}{{Dressler}}{1980}]{Dressler_1980}
{Dressler} A.,  1980, \mn@doi [\apj] {10.1086/157753}, \href {https://ui.adsabs.harvard.edu/abs/1980ApJ...236..351D} {236, 351}

\bibitem[\protect\citeauthoryear{{Driver} et~al.,}{{Driver} et~al.}{2019}]{WAVES}
{Driver} S.~P.,  et~al., 2019, \mn@doi [The Messenger] {10.18727/0722-6691/5126}, \href {https://ui.adsabs.harvard.edu/abs/2019Msngr.175...46D} {175, 46}

\bibitem[\protect\citeauthoryear{{Duncan} et~al.,}{{Duncan} et~al.}{2023}]{ORCHIDSS}
{Duncan} K.,  et~al., 2023, \mn@doi [The Messenger] {10.18727/0722-6691/5306}, \href {https://ui.adsabs.harvard.edu/abs/2023Msngr.190...25D} {190, 25}

\bibitem[\protect\citeauthoryear{{Dye}}{{Dye}}{2008}]{Dye_2008}
{Dye} S.,  2008, \mn@doi [\mnras] {10.1111/j.1365-2966.2008.13639.x}, \href {https://ui.adsabs.harvard.edu/abs/2008MNRAS.389.1293D} {389, 1293}

\bibitem[\protect\citeauthoryear{{Ellison}, {Catinella}  \& {Cortese}}{{Ellison} et~al.}{2018}]{Ellison2018}
{Ellison} S.~L.,  {Catinella} B.,   {Cortese} L.,  2018, \mn@doi [Monthly Notices of the Royal Astronomical Society] {10.1093/mnras/sty1247}, \href {https://ui.adsabs.harvard.edu/abs/2018MNRAS.478.3447E} {478, 3447}

\bibitem[\protect\citeauthoryear{{Fabian}}{{Fabian}}{2012}]{Fabian_2012}
{Fabian} A.~C.,  2012, \mn@doi [{Annual Review of Astronomy and Astrophysics}] {10.1146/annurev-astro-081811-125521}, \href {https://ui.adsabs.harvard.edu/abs/2012ARA&A..50..455F} {50, 455}

\bibitem[\protect\citeauthoryear{{Finlator} \& {Dav{\'e}}}{{Finlator} \& {Dav{\'e}}}{2008}]{Finlator_2008}
{Finlator} K.,  {Dav{\'e}} R.,  2008, \mn@doi [\mnras] {10.1111/j.1365-2966.2008.12991.x}, \href {https://ui.adsabs.harvard.edu/abs/2008MNRAS.385.2181F} {385, 2181}

\bibitem[\protect\citeauthoryear{{Fraser-McKelvie} et~al.,}{{Fraser-McKelvie} et~al.}{2021}]{Fraser-McKelvie_2021}
{Fraser-McKelvie} A.,  et~al., 2021, \mn@doi [\mnras] {10.1093/mnras/stab573}, \href {https://ui.adsabs.harvard.edu/abs/2021MNRAS.503.4992F} {503, 4992}

\bibitem[\protect\citeauthoryear{{Gallazzi} \& {Bell}}{{Gallazzi} \& {Bell}}{2009}]{Gallazzi_2009}
{Gallazzi} A.,  {Bell} E.~F.,  2009, \mn@doi [The Astrophysical Journal, Supplement] {10.1088/0067-0049/185/2/253}, \href {https://ui.adsabs.harvard.edu/abs/2009ApJS..185..253G} {185, 253}

\bibitem[\protect\citeauthoryear{{Geha}, {Blanton}, {Masjedi}  \& {West}}{{Geha} et~al.}{2006}]{Geha_2006}
{Geha} M.,  {Blanton} M.~R.,  {Masjedi} M.,   {West} A.~A.,  2006, \mn@doi [\apj] {10.1086/508604}, \href {https://ui.adsabs.harvard.edu/abs/2006ApJ...653..240G} {653, 240}

\bibitem[\protect\citeauthoryear{Glasser \& Winter}{Glasser \& Winter}{1961}]{src}
Glasser G.~J.,  Winter R.~F.,  1961, Biometrika, 48, 444

\bibitem[\protect\citeauthoryear{{Hale} et~al.,}{{Hale} et~al.}{2024}]{Hale2024}
{Hale} C.~L.,  et~al., 2024, \mn@doi [\mnras] {10.1093/mnras/stae2528}, \href {https://ui.adsabs.harvard.edu/abs/2024MNRAS.tmp.2456H} {}

\bibitem[\protect\citeauthoryear{{Hanish} et~al.,}{{Hanish} et~al.}{2006}]{Hanish_2006}
{Hanish} D.~J.,  et~al., 2006, \mn@doi [\apj] {10.1086/504681}, \href {https://ui.adsabs.harvard.edu/abs/2006ApJ...649..150H} {649, 150}

\bibitem[\protect\citeauthoryear{{Helmboldt}, {Walterbos}, {Bothun}, {O'Neil}  \& {de Blok}}{{Helmboldt} et~al.}{2004}]{Helmboldt_2004}
{Helmboldt} J.~F.,  {Walterbos} R.~A.~M.,  {Bothun} G.~D.,  {O'Neil} K.,   {de Blok} W.~J.~G.,  2004, \mn@doi [\apj] {10.1086/423126}, \href {https://ui.adsabs.harvard.edu/abs/2004ApJ...613..914H} {613, 914}

\bibitem[\protect\citeauthoryear{Heywood et~al.,}{Heywood et~al.}{2021}]{Heywood2021}
Heywood I.,  et~al., 2021, \mn@doi [Monthly Notices of the Royal Astronomical Society] {10.1093/mnras/stab3021}, 509, 2150

\bibitem[\protect\citeauthoryear{{Heywood} et~al.,}{{Heywood} et~al.}{2022}]{Heywood_2022}
{Heywood} I.,  et~al., 2022, \mn@doi [\mnras] {10.1093/mnras/stab3021}, \href {https://ui.adsabs.harvard.edu/abs/2022MNRAS.509.2150H} {509, 2150}

\bibitem[\protect\citeauthoryear{{Heywood} et~al.,}{{Heywood} et~al.}{2024}]{Heywood_2024}
{Heywood} I.,  et~al., 2024, \mn@doi [Monthly Notices of the Royal Astronomical Society] {10.1093/mnras/stae2081}, \href {https://ui.adsabs.harvard.edu/abs/2024MNRAS.534...76H} {534, 76}

\bibitem[\protect\citeauthoryear{{Hildebrand}}{{Hildebrand}}{1983}]{Hildebrand_1983}
{Hildebrand} R.~H.,  1983, \qjras, \href {https://ui.adsabs.harvard.edu/abs/1983QJRAS..24..267H} {24, 267}

\bibitem[\protect\citeauthoryear{{Hopkins}, {Kere{\v{s}}}, {O{\~n}orbe}, {Faucher-Gigu{\`e}re}, {Quataert}, {Murray}  \& {Bullock}}{{Hopkins} et~al.}{2014}]{Hopkins_2014}
{Hopkins} P.~F.,  {Kere{\v{s}}} D.,  {O{\~n}orbe} J.,  {Faucher-Gigu{\`e}re} C.-A.,  {Quataert} E.,  {Murray} N.,   {Bullock} J.~S.,  2014, \mn@doi [Monthly Notices of the Royal Astronomical Society] {10.1093/mnras/stu1738}, \href {https://ui.adsabs.harvard.edu/abs/2014MNRAS.445..581H} {445, 581}

\bibitem[\protect\citeauthoryear{{Huang}, {Haynes}, {Giovanelli}  \& {Brinchmann}}{{Huang} et~al.}{2012}]{Huang_2012a}
{Huang} S.,  {Haynes} M.~P.,  {Giovanelli} R.,   {Brinchmann} J.,  2012, \mn@doi [\apj] {10.1088/0004-637X/756/2/113}, \href {https://ui.adsabs.harvard.edu/abs/2012ApJ...756..113H} {756, 113}

\bibitem[\protect\citeauthoryear{{Huang} et~al.,}{{Huang} et~al.}{2014}]{Huang_2014}
{Huang} S.,  et~al., 2014, \mn@doi [\apj] {10.1088/0004-637X/793/1/40}, \href {https://ui.adsabs.harvard.edu/abs/2014ApJ...793...40H} {793, 40}

\bibitem[\protect\citeauthoryear{{Hunt} et~al.,}{{Hunt} et~al.}{2019}]{Hunt_2019}
{Hunt} L.~K.,  et~al., 2019, \mn@doi [\aap] {10.1051/0004-6361/201834212}, \href {https://ui.adsabs.harvard.edu/abs/2019A&A...621A..51H} {621, A51}

\bibitem[\protect\citeauthoryear{{Hunt}, {Tortora}, {Ginolfi}  \& {Schneider}}{{Hunt} et~al.}{2020}]{Hunt_2020}
{Hunt} L.~K.,  {Tortora} C.,  {Ginolfi} M.,   {Schneider} R.,  2020, \mn@doi [\aap] {10.1051/0004-6361/202039021}, \href {https://ui.adsabs.harvard.edu/abs/2020A&A...643A.180H} {643, A180}

\bibitem[\protect\citeauthoryear{{Ilbert} et~al.,}{{Ilbert} et~al.}{2006a}]{Ilbert2006}
{Ilbert} O.,  et~al., 2006a, \mn@doi [\aap] {10.1051/0004-6361:20065138}, \href {https://ui.adsabs.harvard.edu/abs/2006A&A...457..841I} {457, 841}

\bibitem[\protect\citeauthoryear{{Ilbert} et~al.,}{{Ilbert} et~al.}{2006b}]{lephare2}
{Ilbert} O.,  et~al., 2006b, \mn@doi [\aap] {10.1051/0004-6361:20065138}, \href {https://ui.adsabs.harvard.edu/abs/2006A&A...457..841I} {457, 841}

\bibitem[\protect\citeauthoryear{{Jarvis} et~al.,}{{Jarvis} et~al.}{2013}]{Jarvis2013}
{Jarvis} M.~J.,  et~al., 2013, \mn@doi [\mnras] {10.1093/mnras/sts118}, \href {https://ui.adsabs.harvard.edu/abs/2013MNRAS.428.1281J} {428, 1281}

\bibitem[\protect\citeauthoryear{{Jarvis} et~al.,}{{Jarvis} et~al.}{2016}]{mightee}
{Jarvis} M.,  et~al., 2016, in MeerKAT Science: On the Pathway to the SKA. p.~6 (\mn@eprint {arXiv} {1709.01901})

\bibitem[\protect\citeauthoryear{{Jaskot}, {Oey}, {Salzer}, {Van Sistine}, {Bell}  \& {Haynes}}{{Jaskot} et~al.}{2015}]{Jaskot_2015}
{Jaskot} A.~E.,  {Oey} M.~S.,  {Salzer} J.~J.,  {Van Sistine} A.,  {Bell} E.~F.,   {Haynes} M.~P.,  2015, \mn@doi [The Astrophysical Journal] {10.1088/0004-637X/808/1/66}, \href {https://ui.adsabs.harvard.edu/abs/2015ApJ...808...66J} {808, 66}

\bibitem[\protect\citeauthoryear{Jeffreys}{Jeffreys}{1998}]{jeffreys_1998}
Jeffreys H.,  1998, The theory of probability.
OuP Oxford

\bibitem[\protect\citeauthoryear{Johnson, Wichern  et~al.}{Johnson et~al.}{2002}]{Johnson_2002}
Johnson R.~A.,  Wichern D.~W.,   et~al., 2002, Applied multivariate statistical analysis.
Prentice hall Upper Saddle River, NJ

\bibitem[\protect\citeauthoryear{{Johnston}, {Vaccari}, {Jarvis}, {Smith}, {Giovannoli}, {H{\"a}u{\ss}ler}  \& {Prescott}}{{Johnston} et~al.}{2015}]{johnston2015}
{Johnston} R.,  {Vaccari} M.,  {Jarvis} M.,  {Smith} M.,  {Giovannoli} E.,  {H{\"a}u{\ss}ler} B.,   {Prescott} M.,  2015, \mn@doi [\mnras] {10.1093/mnras/stv1715}, \href {https://ui.adsabs.harvard.edu/abs/2015MNRAS.453.2540J} {453, 2540}

\bibitem[\protect\citeauthoryear{Jonas}{Jonas}{2009}]{meerkat}
Jonas J.~L.,  2009, \mn@doi [Proceedings of the IEEE] {10.1109/JPROC.2009.2020713}, 97, 1522

\bibitem[\protect\citeauthoryear{{Jones} et~al.,}{{Jones} et~al.}{2018}]{Jones_2018}
{Jones} M.~G.,  et~al., 2018, \mn@doi [\aap] {10.1051/0004-6361/201731448}, \href {https://ui.adsabs.harvard.edu/abs/2018A&A...609A..17J} {609, A17}

\bibitem[\protect\citeauthoryear{{Kauffmann} et~al.,}{{Kauffmann} et~al.}{2003}]{Kauffmann_2003}
{Kauffmann} G.,  et~al., 2003, \mn@doi [Monthly Notices of the Royal Astronomical Society] {10.1046/j.1365-8711.2003.06292.x}, \href {https://ui.adsabs.harvard.edu/abs/2003MNRAS.341...54K} {341, 54}

\bibitem[\protect\citeauthoryear{Kendall}{Kendall}{1938}]{ktau}
Kendall M.~G.,  1938, \mn@doi [Biometrika] {10.1093/biomet/30.1-2.81}, 30, 81

\bibitem[\protect\citeauthoryear{{Kennicutt}}{{Kennicutt}}{1998}]{Kennicutt_1998}
{Kennicutt} Robert~C. J.,  1998, \mn@doi [\araa] {10.1146/annurev.astro.36.1.189}, \href {https://ui.adsabs.harvard.edu/abs/1998ARA&A..36..189K} {36, 189}

\bibitem[\protect\citeauthoryear{{Kennicutt} \& {Evans}}{{Kennicutt} \& {Evans}}{2012}]{Kennicutt_2012}
{Kennicutt} R.~C.,  {Evans} N.~J.,  2012, \mn@doi [\araa] {10.1146/annurev-astro-081811-125610}, \href {https://ui.adsabs.harvard.edu/abs/2012ARA&A..50..531K} {50, 531}

\bibitem[\protect\citeauthoryear{{King}}{{King}}{2005}]{King_2005}
{King} A.,  2005, \mn@doi [The Astrophysical Journal, Letters] {10.1086/499430}, \href {https://ui.adsabs.harvard.edu/abs/2005ApJ...635L.121K} {635, L121}

\bibitem[\protect\citeauthoryear{{Kleiner}, {Pimbblet}, {Jones}, {Koribalski}  \& {Serra}}{{Kleiner} et~al.}{2017}]{kleiner_2017}
{Kleiner} D.,  {Pimbblet} K.~A.,  {Jones} D.~H.,  {Koribalski} B.~S.,   {Serra} P.,  2017, \mn@doi [\mnras] {10.1093/mnras/stw3328}, \href {https://ui.adsabs.harvard.edu/abs/2017MNRAS.466.4692K} {466, 4692}

\bibitem[\protect\citeauthoryear{Kraljic, Davé  \& Pichon}{Kraljic et~al.}{2020}]{Kraljic_2020}
Kraljic K.,  Davé R.,   Pichon C.,  2020, \mn@doi [Monthly Notices of the Royal Astronomical Society] {10.1093/mnras/staa250}, 493, 362–381

\bibitem[\protect\citeauthoryear{{Lacey} \& {Cole}}{{Lacey} \& {Cole}}{1993}]{Lacey_1993}
{Lacey} C.,  {Cole} S.,  1993, \mn@doi [Monthly Notices of the Royal Astronomical Society] {10.1093/mnras/262.3.627}, \href {https://ui.adsabs.harvard.edu/abs/1993MNRAS.262..627L} {262, 627}

\bibitem[\protect\citeauthoryear{{Leitner}}{{Leitner}}{2012}]{Leitner_2012}
{Leitner} S.~N.,  2012, \mn@doi [\apj] {10.1088/0004-637X/745/2/149}, \href {https://ui.adsabs.harvard.edu/abs/2012ApJ...745..149L} {745, 149}

\bibitem[\protect\citeauthoryear{{Leja}, {Johnson}, {Conroy}, {van Dokkum}  \& {Byler}}{{Leja} et~al.}{2017}]{Leja_2017}
{Leja} J.,  {Johnson} B.~D.,  {Conroy} C.,  {van Dokkum} P.~G.,   {Byler} N.,  2017, \mn@doi [\apj] {10.3847/1538-4357/aa5ffe}, \href {https://ui.adsabs.harvard.edu/abs/2017ApJ...837..170L} {837, 170}

\bibitem[\protect\citeauthoryear{{Leja}, {Carnall}, {Johnson}, {Conroy}  \& {Speagle}}{{Leja} et~al.}{2019}]{Leja_2019}
{Leja} J.,  {Carnall} A.~C.,  {Johnson} B.~D.,  {Conroy} C.,   {Speagle} J.~S.,  2019, \mn@doi [\apj] {10.3847/1538-4357/ab133c}, \href {https://ui.adsabs.harvard.edu/abs/2019ApJ...876....3L} {876, 3}

\bibitem[\protect\citeauthoryear{{Leslie} et~al.,}{{Leslie} et~al.}{2020}]{Leslie_2020}
{Leslie} S.~K.,  et~al., 2020, \mn@doi [\apj] {10.3847/1538-4357/aba044}, \href {https://ui.adsabs.harvard.edu/abs/2020ApJ...899...58L} {899, 58}

\bibitem[\protect\citeauthoryear{{Lilly}, {Carollo}, {Pipino}, {Renzini}  \& {Peng}}{{Lilly} et~al.}{2013}]{Lilly_2013}
{Lilly} S.~J.,  {Carollo} C.~M.,  {Pipino} A.,  {Renzini} A.,   {Peng} Y.,  2013, \mn@doi [\apj] {10.1088/0004-637X/772/2/119}, \href {https://ui.adsabs.harvard.edu/abs/2013ApJ...772..119L} {772, 119}

\bibitem[\protect\citeauthoryear{{Lin} et~al.,}{{Lin} et~al.}{2010}]{Lin_2010}
{Lin} L.,  et~al., 2010, \mn@doi [The Astrophysical Journal] {10.1088/0004-637X/718/2/1158}, \href {https://ui.adsabs.harvard.edu/abs/2010ApJ...718.1158L} {718, 1158}

\bibitem[\protect\citeauthoryear{{Lonsdale} et~al.,}{{Lonsdale} et~al.}{2003}]{Lonsdale2003}
{Lonsdale} C.~J.,  et~al., 2003, \mn@doi [\pasp] {10.1086/376850}, \href {https://ui.adsabs.harvard.edu/abs/2003PASP..115..897L} {115, 897}

\bibitem[\protect\citeauthoryear{{Lutz} et~al.,}{{Lutz} et~al.}{2017}]{Lutz_2017}
{Lutz} K.~A.,  et~al., 2017, \mn@doi [Monthly Notices of the Royal Astronomical Society] {10.1093/mnras/stx053}, \href {https://ui.adsabs.harvard.edu/abs/2017MNRAS.467.1083L} {467, 1083}

\bibitem[\protect\citeauthoryear{{Macklin}}{{Macklin}}{1982}]{Macklin_1982}
{Macklin} J.~T.,  1982, \mn@doi [\mnras] {10.1093/mnras/199.4.1119}, \href {https://ui.adsabs.harvard.edu/abs/1982MNRAS.199.1119M} {199, 1119}

\bibitem[\protect\citeauthoryear{{Maddox}, {Hess}, {Obreschkow}, {Jarvis}  \& {Blyth}}{{Maddox} et~al.}{2015}]{Maddox_2015}
{Maddox} N.,  {Hess} K.~M.,  {Obreschkow} D.,  {Jarvis} M.~J.,   {Blyth} S.~L.,  2015, \mn@doi [\mnras] {10.1093/mnras/stu2532}, \href {https://ui.adsabs.harvard.edu/abs/2015MNRAS.447.1610M} {447, 1610}

\bibitem[\protect\citeauthoryear{Maddox et~al.,}{Maddox et~al.}{2021}]{Maddox_2021}
Maddox N.,  et~al., 2021, \mn@doi [Astronomy & Astrophysics] {10.1051/0004-6361/202039655}, 646, A35

\bibitem[\protect\citeauthoryear{{Martin}, {Giovanelli}, {Haynes}  \& {Guzzo}}{{Martin} et~al.}{2012}]{Martin_2012}
{Martin} A.~M.,  {Giovanelli} R.,  {Haynes} M.~P.,   {Guzzo} L.,  2012, \mn@doi [\apj] {10.1088/0004-637X/750/1/38}, \href {https://ui.adsabs.harvard.edu/abs/2012ApJ...750...38M} {750, 38}

\bibitem[\protect\citeauthoryear{{Matthee} \& {Schaye}}{{Matthee} \& {Schaye}}{2019}]{Matthee_2019}
{Matthee} J.,  {Schaye} J.,  2019, \mn@doi [Monthly Notices of the Royal Astronomical Society] {10.1093/mnras/stz030}, \href {https://ui.adsabs.harvard.edu/abs/2019MNRAS.484..915M} {484, 915}

\bibitem[\protect\citeauthoryear{{Mauduit} et~al.,}{{Mauduit} et~al.}{2012}]{Mauduit2012}
{Mauduit} J.~C.,  et~al., 2012, \mn@doi [\pasp] {10.1086/666945}, \href {https://ui.adsabs.harvard.edu/abs/2012PASP..124..714M} {124, 714}

\bibitem[\protect\citeauthoryear{{McCracken} et~al.,}{{McCracken} et~al.}{2012}]{McCracken2012}
{McCracken} H.~J.,  et~al., 2012, \mn@doi [\aap] {10.1051/0004-6361/201219507}, \href {https://ui.adsabs.harvard.edu/abs/2012A&A...544A.156M} {544, A156}

\bibitem[\protect\citeauthoryear{{McKee} \& {Ostriker}}{{McKee} \& {Ostriker}}{2007}]{McKee_2007}
{McKee} C.~F.,  {Ostriker} E.~C.,  2007, \mn@doi [\araa] {10.1146/annurev.astro.45.051806.110602}, \href {https://ui.adsabs.harvard.edu/abs/2007ARA&A..45..565M} {45, 565}

\bibitem[\protect\citeauthoryear{{McMullin}, {Waters}, {Schiebel}, {Young}  \& {Golap}}{{McMullin} et~al.}{2007}]{casa}
{McMullin} J.~P.,  {Waters} B.,  {Schiebel} D.,  {Young} W.,   {Golap} K.,  2007, in {Shaw} R.~A.,  {Hill} F.,   {Bell} D.~J.,  eds,  Astronomical Society of the Pacific Conference Series Vol. 376, Astronomical Data Analysis Software and Systems XVI. p.~127

\bibitem[\protect\citeauthoryear{{Meurer} et~al.,}{{Meurer} et~al.}{2006}]{Meurer_2006}
{Meurer} G.~R.,  et~al., 2006, \mn@doi [\apjs] {10.1086/504685}, \href {https://ui.adsabs.harvard.edu/abs/2006ApJS..165..307M} {165, 307}

\bibitem[\protect\citeauthoryear{{Meyer}, {Zwaan}, {Webster}, {Brown}  \& {Staveley-Smith}}{{Meyer} et~al.}{2007}]{Meyer_2007}
{Meyer} M.~J.,  {Zwaan} M.~A.,  {Webster} R.~L.,  {Brown} M. J.~I.,   {Staveley-Smith} L.,  2007, \mn@doi [\apj] {10.1086/508799}, \href {https://ui.adsabs.harvard.edu/abs/2007ApJ...654..702M} {654, 702}

\bibitem[\protect\citeauthoryear{{Mitchell}, {Schaye}, {Bower}  \& {Crain}}{{Mitchell} et~al.}{2020}]{Mitchell_2020}
{Mitchell} P.~D.,  {Schaye} J.,  {Bower} R.~G.,   {Crain} R.~A.,  2020, \mn@doi [\mnras] {10.1093/mnras/staa938}, \href {https://ui.adsabs.harvard.edu/abs/2020MNRAS.494.3971M} {494, 3971}

\bibitem[\protect\citeauthoryear{{Moreno} et~al.,}{{Moreno} et~al.}{2019}]{Moreno_2019}
{Moreno} J.,  et~al., 2019, \mn@doi [\mnras] {10.1093/mnras/stz417}, \href {https://ui.adsabs.harvard.edu/abs/2019MNRAS.485.1320M} {485, 1320}

\bibitem[\protect\citeauthoryear{{Mu{\~n}oz} \& {Peeples}}{{Mu{\~n}oz} \& {Peeples}}{2015}]{Munoz_2015}
{Mu{\~n}oz} J.~A.,  {Peeples} M.~S.,  2015, \mn@doi [\mnras] {10.1093/mnras/stv048}, \href {https://ui.adsabs.harvard.edu/abs/2015MNRAS.448.1430M} {448, 1430}

\bibitem[\protect\citeauthoryear{{Nelson} et~al.,}{{Nelson} et~al.}{2019}]{Nelson_2019}
{Nelson} D.,  et~al., 2019, \mn@doi [\mnras] {10.1093/mnras/stz2306}, \href {https://ui.adsabs.harvard.edu/abs/2019MNRAS.490.3234N} {490, 3234}

\bibitem[\protect\citeauthoryear{{Nersesian} et~al.,}{{Nersesian} et~al.}{2024}]{Nersesian_2024}
{Nersesian} A.,  et~al., 2024, \mn@doi [Astronomy \& Astrophysics] {10.1051/0004-6361/202346769}, \href {https://ui.adsabs.harvard.edu/abs/2024A&A...681A..94N} {681, A94}

\bibitem[\protect\citeauthoryear{{Noeske} et~al.,}{{Noeske} et~al.}{2007}]{Noeske_2007}
{Noeske} K.~G.,  et~al., 2007, \mn@doi [\apjl] {10.1086/517926}, \href {https://ui.adsabs.harvard.edu/abs/2007ApJ...660L..43N} {660, L43}

\bibitem[\protect\citeauthoryear{{Ocvirk}, {Pichon}, {Lan{\c{c}}on}  \& {Thi{\'e}baut}}{{Ocvirk} et~al.}{2006}]{Ocvirk_2006}
{Ocvirk} P.,  {Pichon} C.,  {Lan{\c{c}}on} A.,   {Thi{\'e}baut} E.,  2006, \mn@doi [\mnras] {10.1111/j.1365-2966.2005.09182.x}, \href {https://ui.adsabs.harvard.edu/abs/2006MNRAS.365...46O} {365, 46}

\bibitem[\protect\citeauthoryear{{Oliver} et~al.,}{{Oliver} et~al.}{2012}]{Oliver2012}
{Oliver} S.~J.,  et~al., 2012, \mn@doi [\mnras] {10.1111/j.1365-2966.2012.20912.x}, \href {https://ui.adsabs.harvard.edu/abs/2012MNRAS.424.1614O} {424, 1614}

\bibitem[\protect\citeauthoryear{{Pacifici} et~al.,}{{Pacifici} et~al.}{2023}]{Pacifici_2023}
{Pacifici} C.,  et~al., 2023, \mn@doi [\apj] {10.3847/1538-4357/acacff}, \href {https://ui.adsabs.harvard.edu/abs/2023ApJ...944..141P} {944, 141}

\bibitem[\protect\citeauthoryear{{Pan} et~al.,}{{Pan} et~al.}{2023}]{Pan_2023}
{Pan} H.,  et~al., 2023, \mn@doi [\mnras] {10.1093/mnras/stad2343}, \href {https://ui.adsabs.harvard.edu/abs/2023MNRAS.525..256P} {525, 256}

\bibitem[\protect\citeauthoryear{{Pandya} et~al.,}{{Pandya} et~al.}{2021}]{Pandya_2021}
{Pandya} V.,  et~al., 2021, \mn@doi [\mnras] {10.1093/mnras/stab2714}, \href {https://ui.adsabs.harvard.edu/abs/2021MNRAS.508.2979P} {508, 2979}

\bibitem[\protect\citeauthoryear{{Parkash}, {Brown}, {Jarrett}  \& {Bonne}}{{Parkash} et~al.}{2018}]{Parkash_2018}
{Parkash} V.,  {Brown} M. J.~I.,  {Jarrett} T.~H.,   {Bonne} N.~J.,  2018, \mn@doi [The Astrophysical Journal] {10.3847/1538-4357/aad3b9}, \href {https://ui.adsabs.harvard.edu/abs/2018ApJ...864...40P} {864, 40}

\bibitem[\protect\citeauthoryear{{Peacock}}{{Peacock}}{1983}]{Peacock1983}
{Peacock} J.~A.,  1983, \mn@doi [\mnras] {10.1093/mnras/202.3.615}, \href {https://ui.adsabs.harvard.edu/abs/1983MNRAS.202..615P} {202, 615}

\bibitem[\protect\citeauthoryear{{Peng} et~al.,}{{Peng} et~al.}{2010}]{Peng_2010}
{Peng} Y.-j.,  et~al., 2010, \mn@doi [\apj] {10.1088/0004-637X/721/1/193}, \href {https://ui.adsabs.harvard.edu/abs/2010ApJ...721..193P} {721, 193}

\bibitem[\protect\citeauthoryear{{Peng}, {Lilly}, {Renzini}  \& {Carollo}}{{Peng} et~al.}{2012}]{Peng_2012}
{Peng} Y.-J.,  {Lilly} S.~J.,  {Renzini} A.,   {Carollo} M.,  2012, \mn@doi [\apj] {10.1088/0004-637X/757/1/4}, \href {https://ui.adsabs.harvard.edu/abs/2012ApJ...757....4P} {757, 4}

\bibitem[\protect\citeauthoryear{{Pereira-Wilson}, {Navarro}, {Ben{\'\i}tez-Llambay}  \& {Santos-Santos}}{{Pereira-Wilson} et~al.}{2023}]{Pereira-Wilson_2023}
{Pereira-Wilson} M.,  {Navarro} J.~F.,  {Ben{\'\i}tez-Llambay} A.,   {Santos-Santos} I.,  2023, \mn@doi [\mnras] {10.1093/mnras/stac3633}, \href {https://ui.adsabs.harvard.edu/abs/2023MNRAS.519.1425P} {519, 1425}

\bibitem[\protect\citeauthoryear{{Ponomareva} et~al.,}{{Ponomareva} et~al.}{2023}]{Ponomareva2023}
{Ponomareva} A.~A.,  et~al., 2023, \mn@doi [\mnras] {10.1093/mnras/stad1249}, \href {https://ui.adsabs.harvard.edu/abs/2023MNRAS.522.5308P} {522, 5308}

\bibitem[\protect\citeauthoryear{{Popesso} et~al.,}{{Popesso} et~al.}{2019a}]{Popesso_2019a}
{Popesso} P.,  et~al., 2019a, \mn@doi [\mnras] {10.1093/mnras/sty3210}, \href {https://ui.adsabs.harvard.edu/abs/2019MNRAS.483.3213P} {483, 3213}

\bibitem[\protect\citeauthoryear{{Popesso} et~al.,}{{Popesso} et~al.}{2019b}]{Popesso_2019b}
{Popesso} P.,  et~al., 2019b, \mn@doi [\mnras] {10.1093/mnras/stz2635}, \href {https://ui.adsabs.harvard.edu/abs/2019MNRAS.490.5285P} {490, 5285}

\bibitem[\protect\citeauthoryear{{Rajohnson} et~al.,}{{Rajohnson} et~al.}{2022}]{Rajohnson_2022}
{Rajohnson} S. H.~A.,  et~al., 2022, \mn@doi [\mnras] {10.1093/mnras/stac693}, \href {https://ui.adsabs.harvard.edu/abs/2022MNRAS.512.2697R} {512, 2697}

\bibitem[\protect\citeauthoryear{{Renaud}, {Bournaud}, {Kraljic}  \& {Duc}}{{Renaud} et~al.}{2014}]{Renaud_2014}
{Renaud} F.,  {Bournaud} F.,  {Kraljic} K.,   {Duc} P.~A.,  2014, \mn@doi [\mnras] {10.1093/mnrasl/slu050}, \href {https://ui.adsabs.harvard.edu/abs/2014MNRAS.442L..33R} {442, L33}

\bibitem[\protect\citeauthoryear{{Robotham} et~al.,}{{Robotham} et~al.}{2013}]{Robotham_2013}
{Robotham} A.~S.~G.,  et~al., 2013, \mn@doi [\mnras] {10.1093/mnras/stt156}, \href {https://ui.adsabs.harvard.edu/abs/2013MNRAS.431..167R} {431, 167}

\bibitem[\protect\citeauthoryear{{Robotham} et~al.,}{{Robotham} et~al.}{2014}]{Robotham_2014}
{Robotham} A.~S.~G.,  et~al., 2014, \mn@doi [\mnras] {10.1093/mnras/stu1604}, \href {https://ui.adsabs.harvard.edu/abs/2014MNRAS.444.3986R} {444, 3986}

\bibitem[\protect\citeauthoryear{{Robotham}, {Davies}, {Driver}, {Koushan}, {Taranu}, {Casura}  \& {Liske}}{{Robotham} et~al.}{2018}]{Robotham2018}
{Robotham} A.~S.~G.,  {Davies} L.~J.~M.,  {Driver} S.~P.,  {Koushan} S.,  {Taranu} D.~S.,  {Casura} S.,   {Liske} J.,  2018, \mn@doi [\mnras] {10.1093/mnras/sty440}, \href {https://ui.adsabs.harvard.edu/abs/2018MNRAS.476.3137R} {476, 3137}

\bibitem[\protect\citeauthoryear{{Robotham}, {Bellstedt}, {Lagos}, {Thorne}, {Davies}, {Driver}  \& {Bravo}}{{Robotham} et~al.}{2020}]{ProSPECT}
{Robotham} A.~S.~G.,  {Bellstedt} S.,  {Lagos} C. d.~P.,  {Thorne} J.~E.,  {Davies} L.~J.,  {Driver} S.~P.,   {Bravo} M.,  2020, \mn@doi [\mnras] {10.1093/mnras/staa1116}, \href {https://ui.adsabs.harvard.edu/abs/2020MNRAS.495..905R} {495, 905}

\bibitem[\protect\citeauthoryear{{Rodr{\'\i}guez-Puebla}, {Primack}, {Behroozi}  \& {Faber}}{{Rodr{\'\i}guez-Puebla} et~al.}{2016}]{Rodriguez-Puebla_2016}
{Rodr{\'\i}guez-Puebla} A.,  {Primack} J.~R.,  {Behroozi} P.,   {Faber} S.~M.,  2016, \mn@doi [\mnras] {10.1093/mnras/stv2513}, \href {https://ui.adsabs.harvard.edu/abs/2016MNRAS.455.2592R} {455, 2592}

\bibitem[\protect\citeauthoryear{{Romano} et~al.,}{{Romano} et~al.}{2023}]{Romano_2023}
{Romano} M.,  et~al., 2023, \mn@doi [\aap] {10.1051/0004-6361/202346143}, \href {https://ui.adsabs.harvard.edu/abs/2023A&A...677A..44R} {677, A44}

\bibitem[\protect\citeauthoryear{{Rosenberg}, {Schneider}  \& {Posson-Brown}}{{Rosenberg} et~al.}{2005}]{Rosenberg_2005}
{Rosenberg} J.~L.,  {Schneider} S.~E.,   {Posson-Brown} J.,  2005, \mn@doi [\aj] {10.1086/427397}, \href {https://ui.adsabs.harvard.edu/abs/2005AJ....129.1311R} {129, 1311}

\bibitem[\protect\citeauthoryear{{Saintonge} \& {Catinella}}{{Saintonge} \& {Catinella}}{2022}]{Saintonge_2022}
{Saintonge} A.,  {Catinella} B.,  2022, \mn@doi [\araa] {10.1146/annurev-astro-021022-043545}, \href {https://ui.adsabs.harvard.edu/abs/2022ARA&A..60..319S} {60, 319}

\bibitem[\protect\citeauthoryear{{Saintonge} et~al.,}{{Saintonge} et~al.}{2012}]{Saintonge_2012}
{Saintonge} A.,  et~al., 2012, \mn@doi [The Astrophysical Journal] {10.1088/0004-637X/758/2/73}, \href {https://ui.adsabs.harvard.edu/abs/2012ApJ...758...73S} {758, 73}

\bibitem[\protect\citeauthoryear{{Sanders} et~al.,}{{Sanders} et~al.}{2007}]{Sanders2007}
{Sanders} D.~B.,  et~al., 2007, \mn@doi [\apjs] {10.1086/517885}, \href {https://ui.adsabs.harvard.edu/abs/2007ApJS..172...86S} {172, 86}

\bibitem[\protect\citeauthoryear{{Schiminovich} et~al.,}{{Schiminovich} et~al.}{2010}]{Schiminovich_2010}
{Schiminovich} D.,  et~al., 2010, \mn@doi [\mnras] {10.1111/j.1365-2966.2010.17210.x}, \href {https://ui.adsabs.harvard.edu/abs/2010MNRAS.408..919S} {408, 919}

\bibitem[\protect\citeauthoryear{{Schreiber} et~al.,}{{Schreiber} et~al.}{2018}]{Schreiber_2018}
{Schreiber} C.,  et~al., 2018, \mn@doi [\aap] {10.1051/0004-6361/201833070}, \href {https://ui.adsabs.harvard.edu/abs/2018A&A...618A..85S} {618, A85}

\bibitem[\protect\citeauthoryear{{Serra} et~al.,}{{Serra} et~al.}{2012}]{Serra2012}
{Serra} P.,  et~al., 2012, \mn@doi [\mnras] {10.1111/j.1365-2966.2012.20219.x}, \href {https://ui.adsabs.harvard.edu/abs/2012MNRAS.422.1835S} {422, 1835}

\bibitem[\protect\citeauthoryear{{Silk} \& {Mamon}}{{Silk} \& {Mamon}}{2012}]{Silk_2012}
{Silk} J.,  {Mamon} G.~A.,  2012, \mn@doi [Research in Astronomy and Astrophysics] {10.1088/1674-4527/12/8/004}, \href {https://ui.adsabs.harvard.edu/abs/2012RAA....12..917S} {12, 917}

\bibitem[\protect\citeauthoryear{{Sousbie}}{{Sousbie}}{2011}]{sousbie}
{Sousbie} T.,  2011, \mn@doi [Monthly Notices of the Royal Astronomical Society] {10.1111/j.1365-2966.2011.18394.x}, \href {https://ui.adsabs.harvard.edu/abs/2011MNRAS.414..350S} {414, 350}

\bibitem[\protect\citeauthoryear{{Strateva} et~al.,}{{Strateva} et~al.}{2001}]{Strateva_2001}
{Strateva} I.,  et~al., 2001, \mn@doi [\aj] {10.1086/323301}, \href {https://ui.adsabs.harvard.edu/abs/2001AJ....122.1861S} {122, 1861}

\bibitem[\protect\citeauthoryear{{Tacchella}, {Dekel}, {Carollo}, {Ceverino}, {DeGraf}, {Lapiner}, {Mandelker}  \& {Primack Joel}}{{Tacchella} et~al.}{2016}]{Tacchella_2016}
{Tacchella} S.,  {Dekel} A.,  {Carollo} C.~M.,  {Ceverino} D.,  {DeGraf} C.,  {Lapiner} S.,  {Mandelker} N.,   {Primack Joel} R.,  2016, \mn@doi [\mnras] {10.1093/mnras/stw131}, \href {https://ui.adsabs.harvard.edu/abs/2016MNRAS.457.2790T} {457, 2790}

\bibitem[\protect\citeauthoryear{Tacchella et~al.,}{Tacchella et~al.}{2022}]{Tacchella_2022}
Tacchella S.,  et~al., 2022, \mn@doi [The Astrophysical Journal] {10.3847/1538-4357/ac449b}, 926, 134

\bibitem[\protect\citeauthoryear{{Taylor} et~al.,}{{Taylor} et~al.}{2024}]{Taylor2024}
{Taylor} A.~R.,  et~al., 2024, \mn@doi [\mnras] {10.1093/mnras/stae169}, \href {https://ui.adsabs.harvard.edu/abs/2024MNRAS.528.2511T} {528, 2511}

\bibitem[\protect\citeauthoryear{{Thomas}, {Maraston}, {Schawinski}, {Sarzi}  \& {Silk}}{{Thomas} et~al.}{2010}]{Thomas_2010}
{Thomas} D.,  {Maraston} C.,  {Schawinski} K.,  {Sarzi} M.,   {Silk} J.,  2010, \mn@doi [\mnras] {10.1111/j.1365-2966.2010.16427.x}, \href {https://ui.adsabs.harvard.edu/abs/2010MNRAS.404.1775T} {404, 1775}

\bibitem[\protect\citeauthoryear{{Thorne} et~al.,}{{Thorne} et~al.}{2021}]{Thorne_2021}
{Thorne} J.~E.,  et~al., 2021, \mn@doi [\mnras] {10.1093/mnras/stab1294}, \href {https://ui.adsabs.harvard.edu/abs/2021MNRAS.505..540T} {505, 540}

\bibitem[\protect\citeauthoryear{{Tremonti} et~al.,}{{Tremonti} et~al.}{2004}]{Tremonti_2004}
{Tremonti} C.~A.,  et~al., 2004, \mn@doi [\apj] {10.1086/423264}, \href {https://ui.adsabs.harvard.edu/abs/2004ApJ...613..898T} {613, 898}

\bibitem[\protect\citeauthoryear{{Treyer} et~al.,}{{Treyer} et~al.}{2018}]{Treyer_2018}
{Treyer} M.,  et~al., 2018, \mn@doi [\mnras] {10.1093/mnras/sty769}, \href {https://ui.adsabs.harvard.edu/abs/2018MNRAS.477.2684T} {477, 2684}

\bibitem[\protect\citeauthoryear{{Tudorache} et~al.,}{{Tudorache} et~al.}{2022}]{tudorache2022}
{Tudorache} M.~N.,  et~al., 2022, \mn@doi [\mnras] {10.1093/mnras/stac996}, \href {https://ui.adsabs.harvard.edu/abs/2022MNRAS.513.2168T} {513, 2168}

\bibitem[\protect\citeauthoryear{{Tumlinson}, {Peeples}  \& {Werk}}{{Tumlinson} et~al.}{2017}]{Tumlinson_2017}
{Tumlinson} J.,  {Peeples} M.~S.,   {Werk} J.~K.,  2017, \mn@doi [\araa] {10.1146/annurev-astro-091916-055240}, \href {https://ui.adsabs.harvard.edu/abs/2017ARA&A..55..389T} {55, 389}

\bibitem[\protect\citeauthoryear{{Verdes-Montenegro}, {Yun}, {Williams}, {Huchtmeier}, {Del Olmo}  \& {Perea}}{{Verdes-Montenegro} et~al.}{2001}]{Verdes-Montenegro_2001}
{Verdes-Montenegro} L.,  {Yun} M.~S.,  {Williams} B.~A.,  {Huchtmeier} W.~K.,  {Del Olmo} A.,   {Perea} J.,  2001, \mn@doi [\aap] {10.1051/0004-6361:20011127}, \href {https://ui.adsabs.harvard.edu/abs/2001A&A...377..812V} {377, 812}

\bibitem[\protect\citeauthoryear{{Walcher}, {Groves}, {Budav{\'a}ri}  \& {Dale}}{{Walcher} et~al.}{2011}]{Walcher_2011}
{Walcher} J.,  {Groves} B.,  {Budav{\'a}ri} T.,   {Dale} D.,  2011, \mn@doi [\apss] {10.1007/s10509-010-0458-z}, \href {https://ui.adsabs.harvard.edu/abs/2011Ap&SS.331....1W} {331, 1}

\bibitem[\protect\citeauthoryear{{Walker}, {Johnson}, {Gallagher}, {Privon}, {Kepley}, {Whelan}, {Desjardins}  \& {Zabludoff}}{{Walker} et~al.}{2016}]{Walker_2016}
{Walker} L.~M.,  {Johnson} K.~E.,  {Gallagher} S.~C.,  {Privon} G.~C.,  {Kepley} A.~A.,  {Whelan} D.~G.,  {Desjardins} T.~D.,   {Zabludoff} A.~I.,  2016, \mn@doi [\aj] {10.3847/0004-6256/151/2/30}, \href {https://ui.adsabs.harvard.edu/abs/2016AJ....151...30W} {151, 30}

\bibitem[\protect\citeauthoryear{{Werk} et~al.,}{{Werk} et~al.}{2014}]{Werk_2014}
{Werk} J.~K.,  et~al., 2014, \mn@doi [\apj] {10.1088/0004-637X/792/1/8}, \href {https://ui.adsabs.harvard.edu/abs/2014ApJ...792....8W} {792, 8}

\bibitem[\protect\citeauthoryear{Whitaker, van Dokkum, Brammer  \& Franx}{Whitaker et~al.}{2012}]{Whitaker_2012}
Whitaker K.~E.,  van Dokkum P.~G.,  Brammer G.,   Franx M.,  2012, \mn@doi [The Astrophysical Journal Letters] {10.1088/2041-8205/754/2/L29}, 754, L29

\bibitem[\protect\citeauthoryear{{Whitaker} et~al.,}{{Whitaker} et~al.}{2014}]{Whitaker_2014}
{Whitaker} K.~E.,  et~al., 2014, \mn@doi [\apj] {10.1088/0004-637X/795/2/104}, \href {https://ui.adsabs.harvard.edu/abs/2014ApJ...795..104W} {795, 104}

\bibitem[\protect\citeauthoryear{{Whitaker} et~al.,}{{Whitaker} et~al.}{2015}]{Whitaker_2015}
{Whitaker} K.~E.,  et~al., 2015, \mn@doi [\apjl] {10.1088/2041-8205/811/1/L12}, \href {https://ui.adsabs.harvard.edu/abs/2015ApJ...811L..12W} {811, L12}

\bibitem[\protect\citeauthoryear{Whittaker}{Whittaker}{2009}]{Whittaker_2009}
Whittaker J.,  2009, Graphical models in applied multivariate statistics.
Wiley Publishing

\bibitem[\protect\citeauthoryear{{Wong}, {Meurer}, {Zheng}, {Heckman}, {Thilker}  \& {Zwaan}}{{Wong} et~al.}{2016}]{Wong_2016}
{Wong} O.~I.,  {Meurer} G.~R.,  {Zheng} Z.,  {Heckman} T.~M.,  {Thilker} D.~A.,   {Zwaan} M.~A.,  2016, \mn@doi [\mnras] {10.1093/mnras/stw993}, \href {https://ui.adsabs.harvard.edu/abs/2016MNRAS.460.1106W} {460, 1106}

\bibitem[\protect\citeauthoryear{{Yun}, {Ho}  \& {Lo}}{{Yun} et~al.}{1994}]{Yun_1994}
{Yun} M.~S.,  {Ho} P.~T.~P.,   {Lo} K.~Y.,  1994, \mn@doi [\nat] {10.1038/372530a0}, \href {https://ui.adsabs.harvard.edu/abs/1994Natur.372..530Y} {372, 530}

\bibitem[\protect\citeauthoryear{{Zamojski} et~al.,}{{Zamojski} et~al.}{2007}]{Zamojski2007}
{Zamojski} M.~A.,  et~al., 2007, \mn@doi [\apjs] {10.1086/516593}, \href {https://ui.adsabs.harvard.edu/abs/2007ApJS..172..468Z} {172, 468}

\bibitem[\protect\citeauthoryear{{Zhou}, {Wu}, {Zhou}  \& {Ma}}{{Zhou} et~al.}{2018}]{Zhou_2018}
{Zhou} Z.,  {Wu} H.,  {Zhou} X.,   {Ma} J.,  2018, \mn@doi [\pasp] {10.1088/1538-3873/aad407}, \href {https://ui.adsabs.harvard.edu/abs/2018PASP..130i4101Z} {130, 094101}

\makeatother
\end{thebibliography}

%%%%%%%%%%%%%%%%%%%%%%%%%%%%%%%%%%%%%%%%%%%%%%%%%%

%%%%%%%%%%%%%%%%% APPENDICES %%%%%%%%%%%%%%%%%%%%%

\appendix
\section{Examples of fits and corner plots}
\label{sec:appendix-a}

\pagebreak 

\begin{figure*}
\begin{subfigure}[b]{\linewidth}
  \centering
  % include first image
    \includegraphics[width=0.74\linewidth]{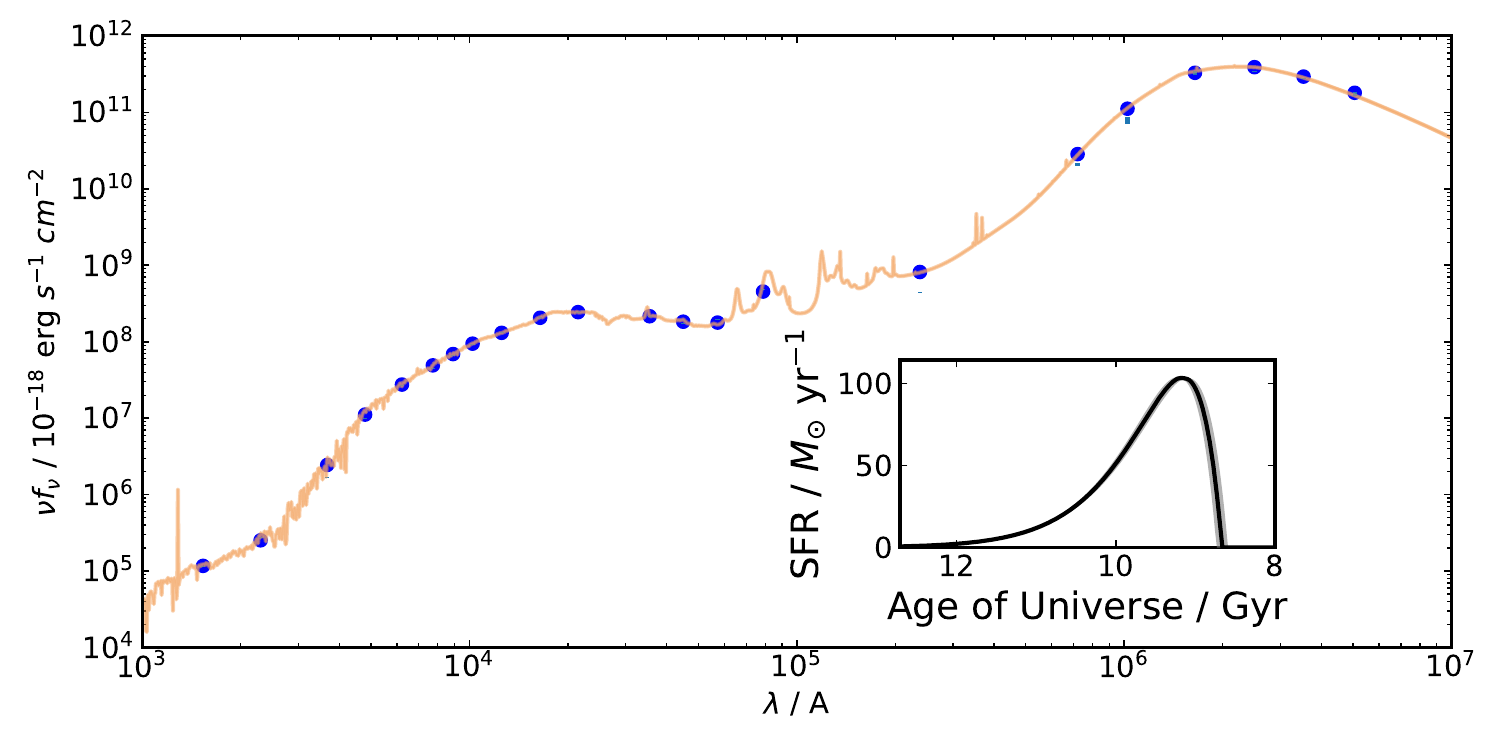}
    \phantomsubcaption
    \label{subfig:photo_sed_82}
\end{subfigure}
\newline
\begin{subfigure}[b]{\linewidth}
  \centering
  % include second image
    \includegraphics[width=0.85\linewidth]{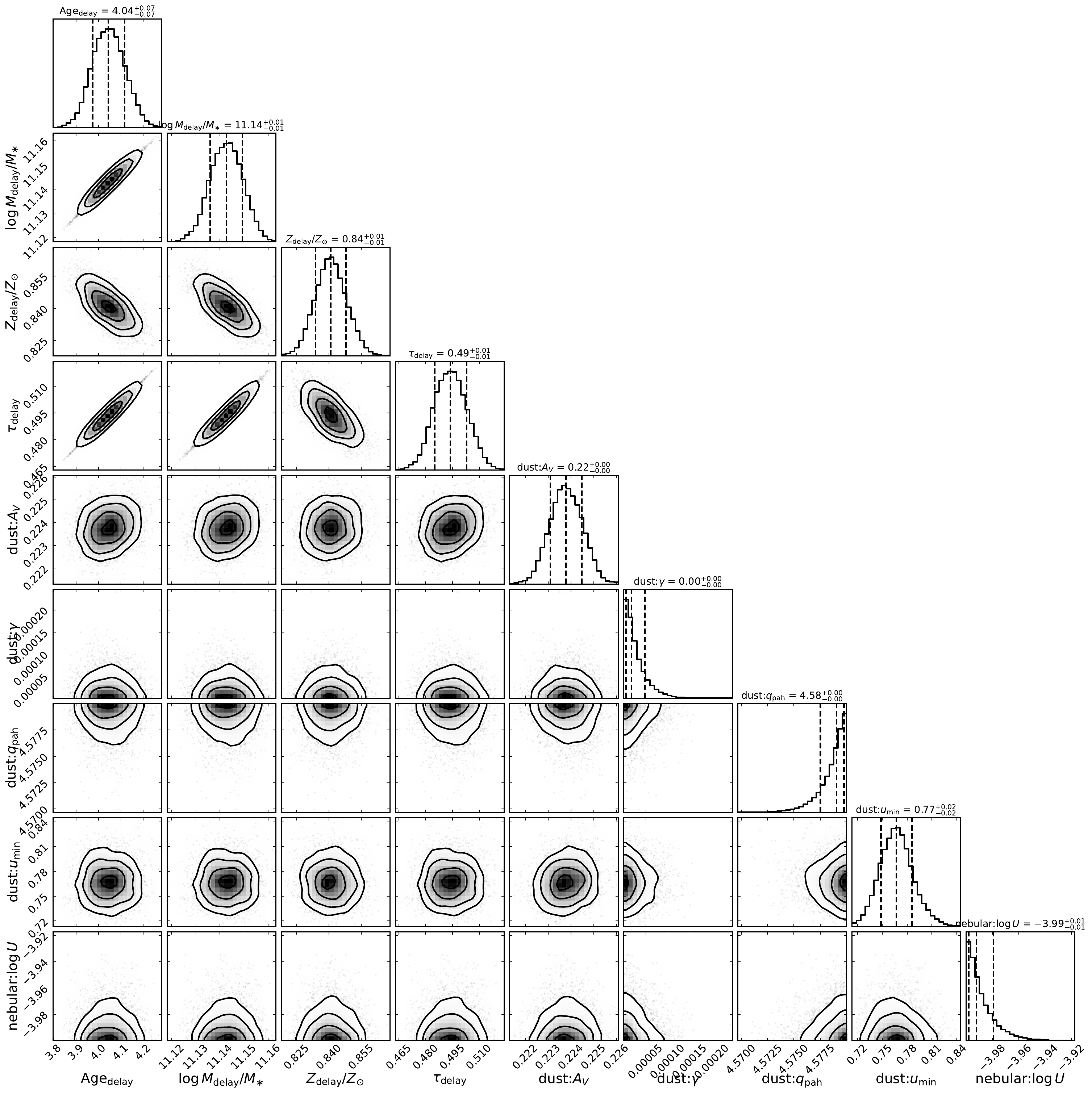}
    \phantomsubcaption
    \label{subfig:corner_82}
\end{subfigure}
\caption{An example of the photometric output of one of the MIGHTEE-H{\sc i} Early Science galaxies using the available photometric filters ({\it top}) and a corner plot ({\it bottom} of one of the MIGHTEE-H{\sc i} Early Science galaxies (Galaxy ID 82, a massive galaxy with $M_{\ast} = 10^{10.86} $~M$_{\sun}$ and SFR of $0.7 \text{M}_{\sun}/\text{yr}$) using the available photometric filters, fitted with an exponentially-delayed SFH model.}
  \label{fig:galaxy_82}
\end{figure*}

\begin{figure*}
\begin{subfigure}[b]{\linewidth}
  \centering
  % include first image
    \includegraphics[width=0.74\linewidth]{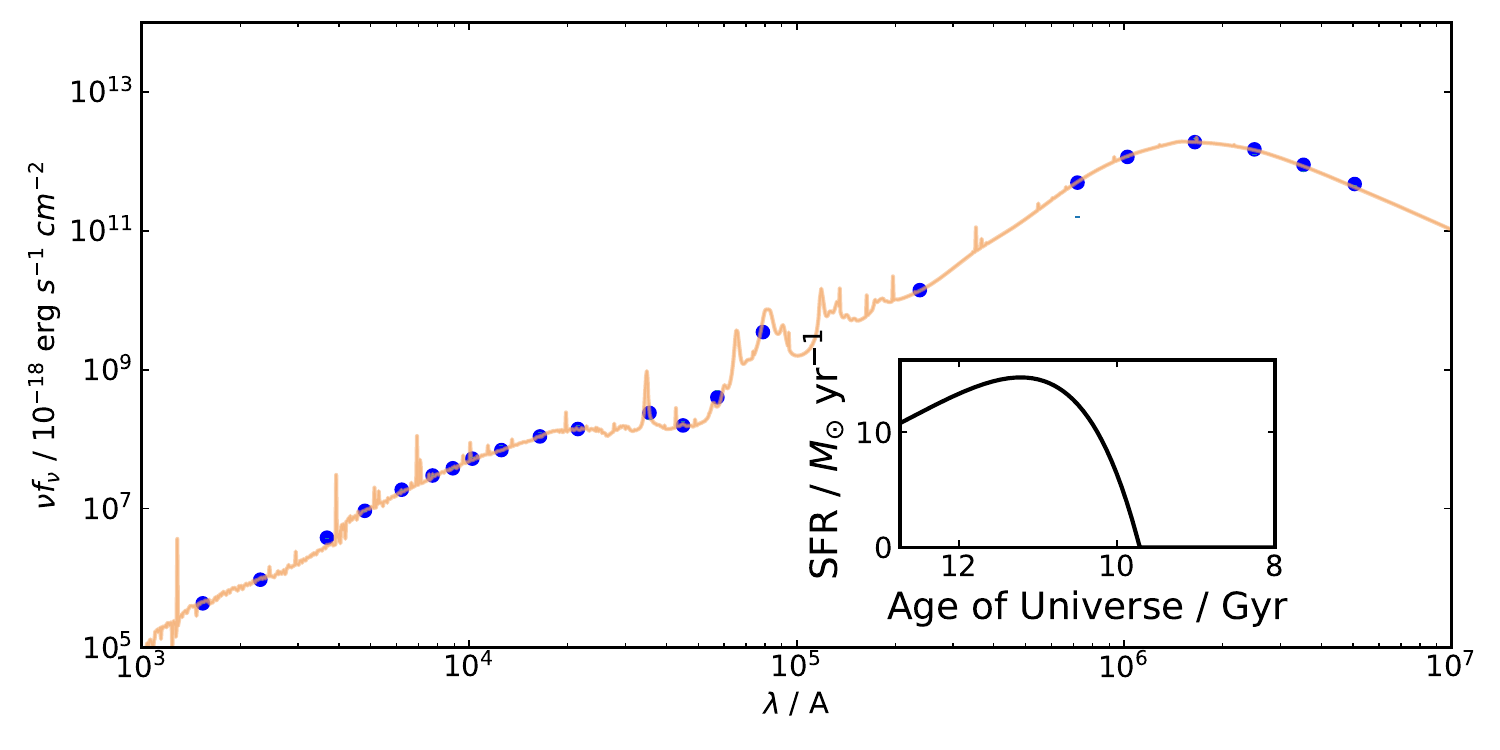}
    \phantomsubcaption
    \label{subfig:photo_sed_158}
\end{subfigure}
\newline
\begin{subfigure}[b]{\linewidth}
  \centering
  % include second image
    \includegraphics[width=0.85\linewidth]{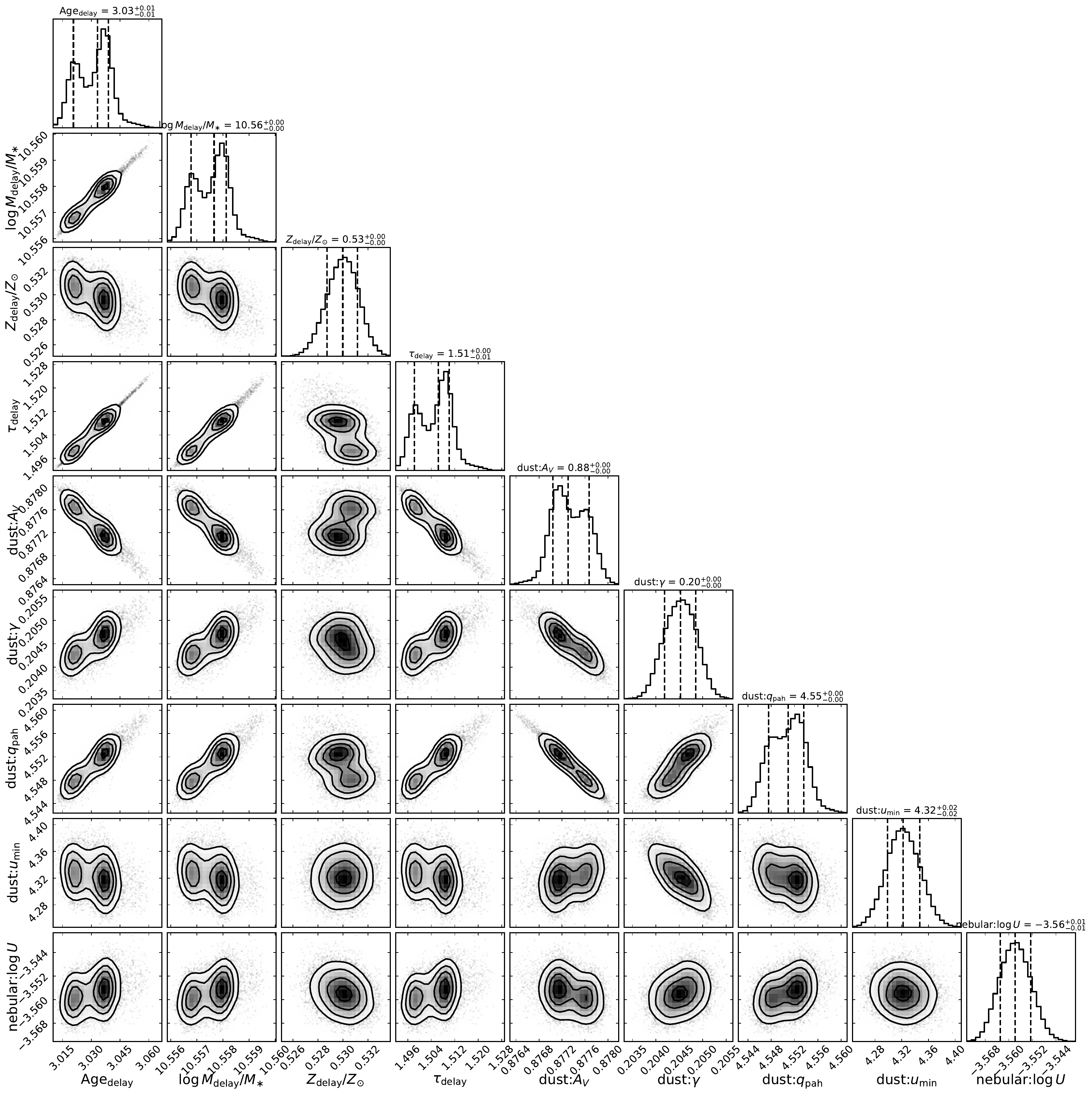}
    \phantomsubcaption
    \label{subfig:corner_158}
\end{subfigure}
\caption{An example of the photometric output of one of the MIGHTEE-H{\sc i} Early Science galaxies using the available photometric filters ({\it top}) and a corner plot ({\it bottom} of one of the MIGHTEE-H{\sc i} Early Science galaxies (Galaxy ID 158, the galaxy with the highest SFR in the sample and $M_{\ast} = 10^{10.28} \text{M}_{\sun}$) using the available photometric filters, fitted with an exponentially-delayed SFH model.}
  \label{fig:galaxy_158}
\end{figure*}

%%%%%%%%%%%%%%%%%%%%%%%%%%%%%%%%%%%%%%%%%%%%%%%%%%

% Don't change these lines
\bsp	% typesetting comment
\label{lastpage}
\end{document}